\documentclass{aa}
\begin{document}

   \thesaurus{04.03.1;  
               05.01.1}  

   \title{Extension of Tycho catalog for low-extinction windows in the galactic bulge}


   \author{T.P. Dominici, R. Teixeira, J.E. Horvath, G.A. Medina Tanco \and P. Benevides-Soares
          }

   \offprints{T.P. Dominici}
  \mail{tania@orion.iagusp.usp.br}

   \institute{Instituto Astronomico e Geof\'\i sico, Universidade de S\~ao Paulo,
              Av. Miguel St\'efano 4200, S\~ao Paulo, SP, 04301-904, Brazil\\
             }

   \date{Received  / accepted }


  \titlerunning{Extension of Tycho catalog}
  \authorrunning{Dominici et al.}

   \maketitle

   \begin{abstract}

   We present in this work secondary catalogs\footnote{Tables B.1 to B.12 are also
available in eletronic form at the CDS via anonymous ftp to
cdsarc.u-strasbg.fr (130.79.128.5) or via
 http://cdsweb.u-strasbg.fr/Abstract.html} up to $m_{Val} \, \sim \, 13$
based on the Tycho reference frame (ESA, 1997) for 12 selected low-extinction fields 
towards the galactic bulge. The observations have been performed with the 
Askania-Zeiss Meridian Circle equiped with a CCD camera, located at the 
Abrah\~ao de Moraes Observatory (Valinhos, Brazil) and operated by the 
Institute of Astronomy and Geophysics, S\~ao Paulo University. The 
presented catalog, though not complete, has been designed to help in 
intensive search programmes (e.g. microlensing and variable searches) 
and therefore the selected standards have a high astrometric and 
photometric ($V$ band, approximately) quality. The mean precisions obtained
were $0.0018^{s}$ in $\alpha$, 0.013'' in $\delta$, 0.030 for the standard
deviation in magnitude and 0.0042 for the magnitude when weighted
with the error bars in each night (in the mean, 42 stars for the catalog
of each window).

   \end{abstract}

%

\section{Introduction}

The large amount of dust and gas in the direction of the galactic bulge is 
known to be responsible for the large extinction which makes very difficult 
the observation of the bulge in several bands. However, it has long been 
known that the irregular distribution of dust and gas presents some "holes" 
where the extinction is considerably lower. The best known of these so-called 
windows has been found by Baade (1951) and extensively studied since then. 
Generally speaking, these low-extinction windows offer a unique opportunity 
to obtain information about the bulge in the optical range. Several groups 
(MACHO, OGLE, for example) have been recently monitoring the bulge in search of 
microlensing events (see Paczy\'nski (1996) for a review).

 Since April 1997 we have been 
conducting a small monitoring project of 12 selected low-extinction windows 
(see Table 1), taken from Blanco \& Terndrup 
(1989) and Blanco (1988), using the recently refurbished Meridian Circle of 
the Abrah\~ao de Moraes Observatory  (see below).
The main objective of the project is the discovery and classification of 
variable stars. The final goal will be to have an on-line, real time 
processing data to stimulate the study of potentially interesting events by 
other observing facilities.

Meanwhile, we have faced the problem that the total number of reference stars of the usual catalogues
inside the observed low-extinction fields (13' in declination and between 3 and 6 minutes
in right ascension) is small. This posed a serious difficulty for 
the reduction of data and the search for variable objects. As a first step 
towards a comprenhensive study of the windows, we have attempted to construct dense
(secondary) reference catalogues intended to be of general use. The publication 
of the Tycho Catalog (ESA, 1997) has added an additional motivation to our task,  
since the refinement of the latter beyond the completeness limit 
($V  \, \sim \, 11$) was already foreseen as an useful tool for future 
observations.

We shall describe in the next Section the selected fields. Section 
III is devoted to a presentation of the instrumental facilities and data 
reduction. The criteria of the selection of the set of standards is 
discussed in Section IV. A general discussion in Section V, the light curves in Appendix A and 
the new catalogues in Appendix B closes the present work. 


\section{Selection of the fields}

Table I summarizes the list of low-extinction windows in which we have defined 
a relatively dense set of standards for further use. Due to specific needs of 
our project, but also in view of more general applications, we have proceeded 
to select standard stars up to $m_{Val} \, = \, 13$ (we will use "$m_{Val}$" to designate the
 magnitudes obtained with the Valinhos system ($V_{V}$ filter) and
"$V$" for magnitudes in the standard Jonhson system ($V_{J}$ filter)). We should emphasize that, 
since the primary objective was to construct an astrometric and photometric reference set, the selected 
stars do {\it not} constitute a {\it complete} sample up to the stated magnitude 
limit. It is also clear that, due to instrumental limitations and sky quality of 
the site, the extension of quality observations to fainter magnitudes is not 
devoided of problems. However, according to the performed tests and preliminary 
results (Dominici, 1998), we are be able to monitor 
 variable stars up to $m_{Val} \, =\, 14.5$; and construct
complete stellar catalogues up to $m_{Val} \, \sim \, 15.5$, the latter
limit being slightly dependent
on the field declination (see below). This relatively 
shallow sampling naturally means that most of our standards do {\it not} reside 
in 
the bulge but are just foreground stars. 
    
\begin{table}[tb]
\caption{Coordinates of the 12 low-extinction windows
for wich secondary catalogs based on Tycho data are presented
(Blanco \& Terndrup, 1989 and  Blanco, 1988) }
\label{tabrmm}
\begin{flushleft}
\begin{tabular}{@{}lcrcrr@{}}
\hline
\rule{0pt}{1.2em}Window&  \# & $\alpha$  (J2000) &$\delta$ (J2000)  &  b ($^{o}$)&l ($^{o}$)\\
\hline
BE  &    5   &    18 10 17    &    -31 45 49    &   -6.02  & 0.20 \\  
BG  &    7   &    18 18 06    &    -32 51 24    &   -7.99  & 0.00  \\
BJ  &    10  &    l8 40 49    &    -34 49 48    &   -13.10 & 0.27  \\
LA  &    11  &    16 56 57    &    -52 44 53    &   -6.02  & -24.77 \\
LB  &    12  &    17 18 27    &    -48 05 40    &   -6.01  & -18.96 \\
LC  &    13  &    17 29 22    &    -45 20 05    &   -6.03  & -15.61 \\
LD  &    14  &    17 44 59    &    -40 41 37    &   -5.99  & -10.14 \\
LI  &    19  &    18 04 48    &    -33 47 46    &   -5.96  & -2.11 \\
LR  &    27  &    18 23 32    &    -26 10 16    &   -5.98  & +6.54\\
LT  &    29  &    18 34 09    &    -21 23 35    &   -6.04  & +9.87 \\
LU  &    30  &    18 51 26    &    -13 18 41    &   -6.03  & +21.04 \\
LV  &    31  &    18 57 37    &    -10 23 35    &   -6.08  & +24.35 \\
\hline
\end{tabular}
\end{flushleft}
\end{table}


\section{Observations and data reduction}

The Askania-Zeiss Abrah\~ao de Moraes Meridian Circle, operated at the IAG/USP 
Valinhos Observatory ($\phi$ = $46^{o}$ 58' 03'', $\lambda$ = $-23^{o}$ 00' 06''), 
is a 0.19 m refractor instrument and 2.6 m focal distance. 
Recently, a CCD detector was installed as a part of the continuing Bordeaux 
Observatory-IAG/USP collaboration (Viateau et al., 1998). 
The capabilities of the Meridian Circle have thus been greatly increased and 
allowed reliable photometric (as well as the 
astrometric) measurements to be performed. 
The CCD detector Thomson 7895A installed at Valinhos has a  
$512 \; \times \, 512$ pixel matrix, with a pixel scale of $1.5"/pixel$. 
It is cooled down to $- 40^{o} \, C$ by a two stage termocouple system. 
The observations are performed in a drift-scanning mode, that is, 
the speed of charge transfer by the CCD is the same as the speed of stellar 
transit for a fixed instrument position. Therefore the integration time for a 
given declination $\delta$ is 

$$ t_{int} \, = \, 51 \, \sec \delta \, s    \, \; \;             (1)$$

The observed field has a width of $13'$ in declination by an arbitrary time 
in right ascention (some minutes or hours). After the installation of the CCD detector, it became 
possible 
to obtain images of several thousands of stars per night, up to magnitudes as 
faint 
as $m_{Val} \, = \, 16.5$, depending on the transit time as reflected by eq.(1).

Although the accuracy of the position and magnitude  
measurements depend on several factors, like the magnitude of 
the observed star, they have been checked to be better than 
0.05 of arcsec for the position and 0.05 for the magnitude (inside the optimal 
magnitude interval $9 \, < \, m_{Val} \, < \, 14$, see Figure 1).

\bigskip


Figure 1:Precision of the positions and magntudes for one typical 
observation night.

\bigskip

The filter we have used in 
this and other observational programmes is somewhat wider than the standard 
Johnson filter $V_{J}$; allowing a larger coverage towards the infrared band 
in order to maximize the number of objects by taking advantage of the better 
quantum efficiency of the CCD in that region. Figure 2 shows the response of 
the Valinhos filter $V_{V}$ together with the standard Johnson filter $V_{J}$, 
both curves already weighted with the CCD detector efficiency.

\bigskip


Figure 2:The response of the Valinhos and Jonhson filters, already weighted with
detector quantum eficiency.

\bigskip

In order to evaluate the difference between our magnitude system and Johnson's
visual band, we selected a filter (kindly provided 
to us by the Laborat\'orio Nacional de Astrof\'\i sica - LNA) capable to roughly
reproducing the necessary spectral characteristics to mimic the $V_{J}$ band.
We used this filter to perform a full observation night, aiming at several
fields for wich we had good data with the broader filter. For the photometric
reduction we made use of the Tycho catalogue $V$ magnitude.

A correlation of the magnitude difference $V \, - \, m_{Val}$ with the 
(poorly determined) color index $(B - V)$ as reported in the Tycho 
catalog is shown in Figure 3 and can be fitted by 

$$ V \, - \, m_{Val} \, = 0.14 \, - \, 0.07 (B - V) \, \, \, (2)$$

However, a large scattering of the data having various origins 
is present and precludes a straightforward utilization of eq.(III.2). 
Moreover, given the intrinsic capabilities of the Meridian Circle we are 
not able to measure $B$ band magnitudes properly because of focusing 
and detector limitations. This means that even though our claims about 
the photometric quality and long-term stability of the selected standards 
are quite secure, those stars should be recalibrated whenever other 
 system is needed for purposes other than differential photometry. 
With this shortcome in mind, an approximate relationship between 
filters may be obtained as 
$m_{Val} \, = \, 1.02 V \, - \, 0.28  $
for quick references purposes only.

\bigskip

Figure 3:Correlation of the magnitude difference ($V - m_{Val}$)
with the color index $(B-V)$

\bigskip

To select a set of standard stars we have performed 5 to 6 observations 
of $\sim \, 1^{h}$ in right ascention each, centered in the selected windows.
 The main objective was to include as 
many Tycho stars as possible in each field to link the selected set to them.
In addition, 10 short observations centered around the field coordinates given 
in Table 1, with a total duration of 3 to 6 minutes each (this is the standard duration of the
observations), were employed to construct light curves for the candidates to 
reference stars and to check the stability of their magnitudes.

The employed data reduction method requires a preliminar 
reduction, 
with the sky background subtracted by a linear polynomial fitted to each pixel 
column. Objects are identified when 3 consecutive pixels with a $2\sigma$ confidence level are detected, where 
$\sigma$ is the standard deviation of the mean count rate in each column. A 
two-dimensional Gaussian curve is adjusted to the flux distribution of the objects, to 
obtain the {\it x} and {\it y} centroid coordinates, the flux and respective errors.
In the following step, the celestial positions and magnitudes are calculated by a global
reduction, using the field overlap among all observation nights (Eichhorn, 1960, Benevides-Soares 
\& Teixeira, 1992, and Teixeira et al., 1992).
In this procedure, first of all, each night is reduced independently of the remaning data by solving the 
following by least squares system, with respect to reference 
catalogue stars:

$$ \alpha \, = \, a_{0} \, + \, a_{1} x \, + \, a_{2} y \; \; 
(3)$$
 
$$ \delta \, = \, b_{0} \, + \, b_{1} y \, + \, b_{2} x  \; \; 
(4)$$
   
$$ m_{Val} \, = \, m_{Val0} \, - \, 2.5 \log F \; \; \; (5) $$
where $a_{0}$ is the initial sidereal time of the observation of the field, $b_{0}$
is the central declination of the field and $x$ and $y$ are the relative coordinates. The $x$
coordinate is measured in the diurnal motion direction while the $y$ coordinate is
in the perpendicular direction. The terms
$a_{1}$ and $b_{1}$ are the scale in $x$ and $y$, respectively, and $a_{2}$ and $b_{2}$ are corrections in
the CCD orientation. A more robust model for the magnitude computation, using color index reported in Tycho catalog, are in 
development, but our tests demonstrated that actual results are confiable. 

After this first reduction, 
the system of eqs.(3-5) is again solved by using an iterative process,
 now for all stars detected in the field. At each step of iteration, the system is solved
by least squares (Benevides-Soares \& Teixeira, 1992 and Teixeira et al., 1992). The process
converges in a few iterations (typically less than 10 steps).

The Tycho Catalog is presented in the form of a multicolumn file in 
which a
quality criterion is given for each object. This criterion guides the further 
utilization of the stars as standards for any purpose. We have concluded that the
construction of secondary catalogues referred to Tycho stars of quality index
worse than 5 can compromised the desired accuracy, and therefore we excluded such Tycho 
stars. Stars brighter than $m_{Val} \, = \, 8.5$ have been 
also excluded because they have a saturated image 
in the CCD and their use would  
spoil the magnitude mesurement of all the stars in the field. All the known 
variable 
stars were also taken out according to the  
GCVS (Khopolov et al, 1988) and NSV 
(Kukarkin  et al, 1982) Catalogues, and the new variables found by the HIPPARCOS 
mission. With this criteria the final number of Tycho stars present in the "short" exposure lies between 1 and 8,
and  in the "long" exposures, between 31 and 89 stars.  


\section{Construction of the Reference Catalogue}

After reduction of the "long" images, as described in the previous Section, we 
have separated and relabeled all the stars brighter than $m_{Val} \, =\, 13$ that 
appeared in all the observational runs. They constitued the first set of 
candidates for a reference catalogue. A visual inspection was performed to 
eliminate those stars located near the edges of the images and those with 
very near apparent neighbours (which may have complicated the magnitude 
determination and introduce errors in position coordinates as well).
The number of "long" observations was certainly enough to guarantee 
the astrometric accuracy of the candidates, but not their photometric 
stability.
Thus, to detect possible variables among our reference candidates,
 we have employed 10 "short" exposure observations (in principle
obtained for the monitoring project) in addition to the 
"long" ones to construct the appropriate light curves.
For a few fields, the "short" observations did not contain
the minimum of four Tycho stars, necessary for the reduction 
procedure. In these cases, the number of reference stars
was increased with candidates stars from the "long" observations.
 On the other 
hand, this procedure allowed us to check the sensitivity of the results to a 
change in the input standard stars.
Once we had constructed all the light curves for the remaining candidates we 
proceeded to eliminate those which had large magnitude and positions variations based on the 
derived parameters: position and magnitude average values, their corresponding 
standard deviations($\sigma_{\alpha}$, $\sigma_{\delta}$, $\sigma_{m_{Val}}$),
 the average computed for the magnitude weighted with 
the error bars in each night and weighted error $\sigma_{w}$. 
This selection eliminated  
stars having 
a magnitude standard deviation larger than 0.1; right ascention 
errors larger than $0.01^{s}$ and declination errors larger than $0.1"$. 
In addition, stars for which the mean value of the magnitude was significantly 
different from the average weighted with the nightly error bars were also 
discarded (see examples of the light curves in Appendix A, Figs.4-6). In Table 3 we present
the average values of the errors in position and magnitude and the number of stars
in the final catalog.

\begin{table}[tb]
\caption{Average position and magnitude  precisions for reference stars in the low-extinction windows. The
last colum indicate the number of stars in the final catalogs}
\label{tabrmm}
\begin{flushleft}
\begin{tabular}{@{}lcrccc@{}}
\hline
\rule{0pt}{1.2em}Window& $\sigma_\alpha(^{s})$& $\sigma_\delta('')$& $\sigma_{m_{Val}}$& $\sigma_{w}$& N\\
\hline
BE     &	0.0017	&	0.015	&	0.028 &		0.0043 &  60\\
BG     &	0.0021	&	0.014	&	0.026 &		0.0040 &  41\\
BJ     &	0.0021	&	0.013	&	0.027 &		0.0041 &  43\\
LA     &	0.0019	&	0.011	&	0.035 &		0.0038 &  55\\
LB     &	0.0019	&	0.011	&	0.033 &		0.0039 &  36\\
LC     &	0.0014	&	0.011	&	0.029 &		0.0035 &  61\\
LD     &	0.0015	&	0.011	&	0.027 &		0.0036 &  47\\
LI     &	0.0010	&	0.011	&	0.031 &		0.0038 &  20\\
LR     &	0.0025	&	0.014	&	0.031 &		0.0048 &  37\\
LT     &	0.0021	&	0.014	&	0.028 &		0.0049 &  32\\
LU     &	0.0017	&	0.013	&	0.028 &		0.0045 &  31\\
LV     &	0.0015	&	0.016	&	0.034 &		0.0048 &  44\\
\hline
\end{tabular}
\end{flushleft}
\end{table}


\section{Concluding remarks}

 The data from standard stars appearing in this work are the result of the best 
quality images obtained in April-September 1997 (approximately 54 nights in total, average of 
30 observations per window), reduced and processed
according to the "global" procedures discussed in Section II. Some stars 
preliminary selected as standards were in fact 
eliminated after the end of 1997 campaign because they did not fully satisfy the
selection criteria in the long run. We note that all these standards are being 
observed daily (whenever observational conditions are satisfactory) as a 
byproduct of our monitoring programme, and
therefore the results of our catalogues are continuously being improved. 
Upgraded versions and the finding charts of the catalogues can be requested electronically.

\begin{acknowledgements}
      We would like to thanks the financial support by FAPESP, CAPES, CNPq, Bordeaux Observatory and
IAG/USP, R.D.D da Costa and the referee, F.Mignard, for useful suggestions that helped to 
improve the first version of the present work. 
\end{acknowledgements}

\appendix
\section{Light curves examples}

Figs.4-6 present some examples of typical light curves accepted for the final 
reference set selection.

\bigskip


Figure 4:Examples of light curves of the catalog stars in the windows BE, BG, BJ e LA.

\bigskip

Figure 5:Examples of light curves of the catalog stars in the windows LB, LC, LD e LI.

\bigskip

Figure 6:Examples of light curves of the catalog stars in the windows LR, LT, LU e LV.

\bigskip

\section{Valinhos Reference Catalogues}

The Tables contain the recommended reference stars for the 12 low-extinction
fields towards the galactic bulge. The first column of the catalog indicates  
the star label, formed by a number that designates the window (see Table
1) and a sequential counter. The following colums contain the 
mean right ascention, the corresponding standard deviation (in seconds); the
mean declination, its standard deviation (in arcseconds) (J2000); the weighted 
average magnitude, its standard deviation and weighted error.
Stars belonging to the 
Tycho Catalogue are indicated with a $T$  
in the first column.
 We stress again 
that, being a natural extension of the Tycho Catalogue, the Valinhos Reference 
Catalogue is {\it not} complete up to the given magnitude, but instead 
constitude a reference set which may prove useful for dedicated studies of 
low-extinction fields. Complete databases up to $m_{Val} \, = 15.5$ or better 
will be available soon as a product of our monitoring programme and will 
be published elsewhere.

\begin{table*}[tb]
\caption{Secondary catalog of references for the BE window (mean epoch 2450612, based on 24 nights
of observations)}
\label{tabrmm}
\begin{flushleft}
\begin{tabular}{@{}lcrcrccc@{}}
\hline
\rule{0pt}{1.2em}Name& $\alpha$ ($^{h}$ $^{m}$ $^{s}$)&$\sigma_{\alpha}$ ($^{s}$)&$\delta$ ($^{o}$ ' '')&$\sigma_{\delta} ('')$&$m_{Val}$&$\sigma_{m_{Val}}$&$\sigma_{w}$\\
\hline
5      272&	18 08 08.3421&	0.0013&	-31 42 34.542&	0.020&	11.822&	0.094&	0.0041\\
5      274&	18 08 14.7501&	0.0009&	-31 48 02.973&	0.019&	12.501&	0.032&	0.0040\\
5      278&	18 08 25.5116&	0.0032&	-31 46 00.176&	0.018&	12.410&	0.035&	0.0037\\
5      279&	18 08 26.9009&	0.0011&	-31 44 43.432&	0.016&	11.694&	0.041&	0.0060\\
5      280&	18 08 29.2174&	0.0008&	-31 44 44.253&	0.017&	11.914&	0.035&	0.0037\\
5      282&	18 08 32.5417&	0.0023&	-31 43 05.793&	0.015&	12.106&	0.037&	0.0046\\
5      286&	18 08 35.2431&	0.0023&	-31 51 11.000&	0.020&	12.095&	0.020&	0.0039\\
5      287&	18 08 39.2616&	0.0007&	-31 44 25.079&	0.013&	12.024&	0.024&	0.0060\\
5      288&	18 08 39.4759&	0.0008&	-31 50 18.990&	0.014&	12.364&	0.019&	0.0038\\
5      289&	18 08 40.1803&	0.0016&	-31 40 27.421&	0.016&	13.004&	0.032&	0.0048\\
5      291&	18 08 42.5087&	0.0007&	-31 41 50.519&	0.013&	11.947&	0.020&	0.0032\\
5      292&	18 08 43.9389&	0.0008&	-31 50 04.194&	0.013&	12.738&	0.027&	0.0050\\
5      293&	18 08 44.4525&	0.0034&	-31 41 17.756&	0.013&	12.763&	0.024&	0.0045\\
5      295&	18 08 47.7469&	0.0030&	-31 42 57.130&	0.019&	12.969&	0.045&	0.0052\\
5      296&	18 08 49.6583&	0.0008&	-31 43 42.182&	0.016&	12.343&	0.024&	0.0044\\
5      298&	18 08 56.9861&	0.0025&	-31 46 56.901&	0.016&	11.644&	0.027&	0.0038\\
5      299&	18 08 59.1021&	0.0015&	-31 43 50.789&	0.013&	11.133&	0.018&	0.0057\\
5      300&	18 09 01.4015&	0.0009&	-31 46 39.589&	0.011&	12.687&	0.024&	0.0044\\
5      301&	18 09 03.5565&	0.0050&	-31 48 45.255&	0.013&	13.068&	0.047&	0.0050\\
5      302&	18 09 03.7020&	0.0024&	-31 49 28.720&	0.016&	11.628&	0.032&	0.0035\\
5      304&	18 09 04.6657&	0.0039&	-31 42 35.194&	0.014&	12.260&	0.023&	0.0042\\
5      305&	18 09 04.8513&	0.0026&	-31 45 04.393&	0.015&	12.636&	0.030&	0.0043\\
5      313&	18 09 17.0028&	0.0026&	-31 45 10.084&	0.013&	11.375&	0.030&	0.0031\\
5      319&	18 09 27.9283&	0.0005&	-31 43 36.101&	0.015&	12.244&	0.037&	0.0048\\
5      321&	18 09 33.9621&	0.0007&	-31 45 21.332&	0.014&	11.097&	0.011&	0.0055\\
5      322&	18 09 37.4655&	0.0013&	-31 47 13.957&	0.015&	11.690&	0.031&	0.0048\\
5      323&	18 09 37.6239&	0.0009&	-31 46 14.679&	0.016&	11.175&	0.030&	0.0033\\
5      324&	18 09 38.1957&	0.0041&	-31 49 40.720&	0.016&	12.213&	0.024&	0.0049\\
5      325&	18 09 41.8092&	0.0008&	-31 45 14.798&	0.014&	12.941&	0.036&	0.0052\\
5      327&	18 09 46.6390&	0.0008&	-31 40 14.103&	0.015&	12.490&	0.038&	0.0042\\
5      328&	18 09 48.0775&	0.0026&	-31 47 08.111&	0.014&	12.398&	0.019&	0.0042\\
5      329&	18 09 50.6644&	0.0012&	-31 44 16.845&	0.017&	12.827&	0.032&	0.0049\\
5      332&	18 09 58.8195&	0.0018&	-31 46 11.992&	0.018&	12.398&	0.027&	0.0040\\
5      334&	18 10 04.5103&	0.0026&	-31 51 00.773&	0.021&	12.256&	0.027&	0.0039\\
5      335&	18 10 05.7402&	0.0026&	-31 43 41.039&	0.018&	11.344&	0.015&	0.0044\\
5      337&	18 10 08.0214&	0.0009&	-31 45 15.133&	0.018&	12.471&	0.021&	0.0040\\
5      339&	18 10 15.5005&	0.0012&	-31 47 07.518&	0.020&	12.768&	0.021&	0.0050\\
5      341&	18 10 23.3952&	0.0011&	-31 44 15.764&	0.014&	12.391&	0.021&	0.0038\\
5      344&	18 10 32.0785&	0.0014&	-31 40 55.609&	0.016&	12.906&	0.030&	0.0048\\
5      346&	18 10 33.9527&	0.0009&	-31 44 06.511&	0.014&	12.798&	0.034&	0.0048\\
5      347&	18 10 34.0943&	0.0010&	-31 47 00.357&	0.015&	12.773&	0.022&	0.0044\\
5      348&	18 10 37.7693&	0.0010&	-31 43 42.808&	0.015&	11.317&	0.021&	0.0023\\
5      349&	18 10 39.1303&	0.0011&	-31 47 34.388&	0.010&	12.198&	0.038&	0.0049\\
5      350 - T&	18 10 40.4400&	0.0014&	-31 47 40.400&	0.012&	10.847&	0.030&	0.0031\\
5      351&	18 10 42.4630&	0.0008&	-31 45 17.907&	0.014&	12.275&	0.024&	0.0039\\
5      353&	18 10 44.5643&	0.0010&	-31 42 23.987&	0.014&	12.071&	0.020&	0.0037\\
5      354&	18 10 47.8467&	0.0008&	-31 45 46.460&	0.012&	12.198&	0.037&	0.0040\\
5      355&	18 10 51.2393&	0.0094&	-31 41 08.018&	0.012&	11.562&	0.021&	0.0027\\
5      357&	18 10 54.8549&	0.0027&	-31 43 06.436&	0.014&	12.212&	0.014&	0.0034\\
5      358&	18 10 55.2818&	0.0011&	-31 48 00.218&	0.012&	11.991&	0.027&	0.0040\\
5      360&	18 11 03.1261&	0.0010&	-31 40 16.847&	0.015&	12.313&	0.040&	0.0044\\
5      363&	18 11 09.1148&	0.0008&	-31 45 24.472&	0.013&	11.709&	0.017&	0.0039\\
5      364&	18 11 13.8770&	0.0009&	-31 51 00.091&	0.016&	12.901&	0.019&	0.0058\\
5      368&	18 11 20.4411&	0.0007&	-31 45 24.474&	0.012&	11.816&	0.018&	0.0040\\
5      372&	18 11 32.6107&	0.0007&	-31 44 40.948&	0.015&	12.143&	0.036&	0.0041\\
5      378&	18 11 57.8431&	0.0005&	-31 40 42.243&	0.012&	11.986&	0.026&	0.0042\\
5      381&	18 12 05.7049&	0.0009&	-31 41 46.573&	0.012&	10.871&	0.021&	0.0040\\
5      384&	18 12 19.8485&	0.0005&	-31 45 09.247&	0.017&	12.890&	0.026&	0.0045\\
5      389&	18 12 31.0789&	0.0010&	-31 40 06.953&	0.014&	11.372&	0.030&	0.0043\\
5      391&	18 12 34.1459&	0.0021&	-31 40 47.207&	0.018&	12.459&	0.032&	0.0039\\
\hline    
\end{tabular}
\end{flushleft}
\end{table*}

\begin{table*}[tb]
\caption{Secondary catalog of references for the BG window (mean epoch 2450617, based on 23 nights
of observations)}
\label{tabrmm}
\begin{flushleft}
\begin{tabular}{@{}lcrcrccc@{}}
\hline
\rule{0pt}{1.2em}Name& $\alpha$ ($^{h}$ $^{m}$ $^{s}$)&$\sigma_{\alpha}$ ($^{s}$)&$\delta$ ($^{o}$ ' '')&$\sigma_{\delta} ('')$&$m_{Val}$&$\sigma_{m_{Val}}$&$\sigma_{w}$\\
\hline
7      302&	18 16 11.6536&	0.0007&	-32 46 44.373&	0.015&	12.783&	0.026&	0.0040\\
7      303&	18 16 14.7920&	0.0009&	-32 51 10.386&	0.016&	12.883&	0.026&	0.0042\\
7      304&	18 16 15.6959&	0.0016&	-32 54 10.196&	0.013&	12.902&	0.046&	0.0046\\
7      305&	18 16 16.9887&	0.0036&	-32 46 33.338&	0.014&	12.813&	0.045&	0.0052\\
7      306&	18 16 18.4427&	0.0013&	-32 46 49.787&	0.016&	13.003&	0.037&	0.0045\\
7      307&	18 16 22.6818&	0.0097&	-32 54 16.040&	0.011&	12.177&	0.039&	0.0045\\
7      310&	18 16 25.4818&	0.0025&	-32 49 43.309&	0.011&	12.007&	0.024&	0.0036\\
7      311&	18 16 26.5953&	0.0011&	-32 47 23.624&	0.011&	12.368&	0.028&	0.0041\\
7      313&	18 16 33.1180&	0.0010&	-32 49 53.235&	0.011&	12.343&	0.023&	0.0037\\
7      314&	18 16 45.7434&	0.0006&	-32 53 00.176&	0.016&	12.425&	0.024&	0.0038\\
7      317&	18 16 55.9462&	0.0091&	-32 53 45.457&	0.017&	12.679&	0.034&	0.0040\\
7      320&	18 17 00.8620&	0.0143&	-32 46 49.281&	0.016&	12.455&	0.030&	0.0040\\
7      323&	18 17 04.4974&	0.0011&	-32 54 56.226&	0.015&	12.176&	0.028&	0.0037\\
7      324&	18 17 10.9413&	0.0009&	-32 48 29.762&	0.012&	12.429&	0.025&	0.0033\\
7      325&	18 17 11.7457&	0.0018&	-32 48 49.947&	0.011&	12.439&	0.021&	0.0038\\
7      328&	18 17 24.4044&	0.0007&	-32 48 02.122&	0.014&	12.877&	0.034&	0.0044\\
7      330&	18 17 28.5190&	0.0015&	-32 46 49.741&	0.011&	12.937&	0.021&	0.0043\\
7      334&	18 17 47.5671&	0.0008&	-32 49 28.578&	0.015&	12.360&	0.025&	0.0032\\
7      336&	18 17 48.8659&	0.0028&	-32 48 21.882&	0.016&	12.634&	0.022&	0.0040\\
7      337&	18 17 51.5368&	0.0052&	-32 48 53.377&	0.019&	11.573&	0.016&	0.0030\\
7      343&	18 18 10.6196&	0.0008&	-32 46 29.046&	0.016&	12.199&	0.036&	0.0042\\
7      345&	18 18 15.5172&	0.0012&	-32 50 22.667&	0.016&	12.702&	0.032&	0.0043\\
7      346&	18 18 19.0545&	0.0015&	-32 48 15.311&	0.016&	13.011&	0.028&	0.0046\\
7      348&	18 18 19.8340&	0.0007&	-32 52 47.977&	0.016&	12.024&	0.024&	0.0038\\
7      351&	18 18 22.8957&	0.0009&	-32 55 28.541&	0.015&	12.216&	0.014&	0.0039\\
7      352&	18 18 22.9885&	0.0013&	-32 53 33.282&	0.014&	11.720&	0.035&	0.0025\\
7      354&	18 18 38.3786&	0.0012&	-32 46 40.386&	0.015&	12.677&	0.021&	0.0040\\
7      355&	18 18 39.7025&	0.0010&	-32 48 29.090&	0.014&	12.596&	0.019&	0.0041\\
7      356 - T&	18 18 40.2599&	0.0006&	-32 49 13.300&	0.013&	10.986&	0.017&	0.0024\\
7      357&	18 18 45.9533&	0.0011&	-32 52 59.369&	0.013&	12.874&	0.026&	0.0047\\
7      358&	18 18 51.5895&	0.0010&	-32 55 48.784&	0.009&	11.414&	0.016&	0.0049\\
7      359 - T&	18 18 53.0100&	0.0010&	-32 51 46.800&	0.011&	11.324&	0.015&	0.0039\\
7      360&	18 18 55.2087&	0.0011&	-32 47 30.083&	0.013&	12.721&	0.023&	0.0042\\
7      361&	18 18 55.4000&	0.0031&	-32 52 12.013&	0.012&	11.059&	0.023&	0.0037\\
7      364&	18 19 04.1083&	0.0010&	-32 53 22.673&	0.013&	12.464&	0.030&	0.0035\\
7      366&	18 19 06.7211&	0.0005&	-32 46 19.106&	0.010&	11.279&	0.030&	0.0039\\
7      368&	18 19 11.3597&	0.0008&	-32 50 13.223&	0.011&	11.744&	0.016&	0.0042\\
7      370&	18 19 19.2782&	0.0008&	-32 56 02.298&	0.015&	12.560&	0.024&	0.0040\\
7      371&	18 19 27.8979&	0.0031&	-32 47 23.341&	0.014&	12.763&	0.027&	0.0041\\
7      373&	18 19 30.0527&	0.0011&	-32 49 32.940&	0.012&	12.482&	0.018&	0.0039\\
7      375&	18 19 39.1025&	0.0010&	-32 53 38.524&	0.011&	12.146&	0.029&	0.0035\\
\hline    
\end{tabular}
\end{flushleft}
\end{table*}

\begin{table*}[tb]
\caption{Secondary catalog of references for the BJ window (mean epoch 2450617, based on 23 nights
of observations)}
\label{tabrmm}
\begin{flushleft}
\begin{tabular}{@{}lcrcrccc@{}}
\hline
\rule{0pt}{1.2em}Name& $\alpha$ ($^{h}$ $^{m}$ $^{s}$)&$\sigma_{\alpha}$ ($^{s}$)&$\delta$ ($^{o}$ ' '')&$\sigma_{\delta} ('')$&$m_{Val}$&$\sigma_{m_{Val}}$&$\sigma_{w}$\\
\hline
10     332&	18 39 31.0121&	0.0087&	-34 44 55.456&	0.017&	12.413&	0.046&	0.0043\\
10     333&	18 39 43.0271&	0.0073&	-34 53 46.619&	0.009&	11.284&	0.014&	0.0050\\
10     334&	18 39 43.6727&	0.0017&	-34 44 14.205&	0.011&	12.444&	0.025&	0.0036\\
10     335&	18 39 45.0843&	0.0013&	-34 44 07.739&	0.010&	12.316&	0.032&	0.0040\\
10     337&	18 39 50.2354&	0.0013&	-34 47 34.385&	0.012&	12.534&	0.018&	0.0036\\
10     339&	18 39 56.7345&	0.0011&	-34 48 00.661&	0.010&	12.853&	0.030&	0.0043\\
10     341 - T&	18 39 58.0100&	0.0006&	-34 44 13.400&	0.008&	10.960&	0.019&	0.0020\\
10     343&	18 40 04.3981&	0.0008&	-34 45 06.438&	0.013&	12.889&	0.021&	0.0044\\
10     344&	18 40 08.9125&	0.0011&	-34 50 37.503&	0.014&	12.107&	0.034&	0.0036\\
10     346&	18 40 14.5296&	0.0009&	-34 51 08.484&	0.016&	12.431&	0.040&	0.0042\\
10     347&	18 40 21.6425&	0.0009&	-34 50 31.361&	0.014&	12.483&	0.047&	0.0040\\
10     348&	18 40 24.2976&	0.0008&	-34 47 39.068&	0.011&	12.324&	0.026&	0.0037\\
10     349&	18 40 25.5534&	0.0157&	-34 46 40.913&	0.015&	12.640&	0.023&	0.0040\\
10     350&	18 40 27.6624&	0.0030&	-34 48 35.587&	0.011&	11.529&	0.024&	0.0039\\
10     351&	18 40 35.0017&	0.0013&	-34 50 24.946&	0.010&	12.082&	0.041&	0.0037\\
10     352 - T&	18 40 35.0700&	0.0022&	-34 45 39.100&	0.009&	10.631&	0.018&	0.0034\\
10     353&	18 40 35.7349&	0.0007&	-34 50 44.776&	0.012&	12.900&	0.041&	0.0044\\
10     354&	18 40 41.0146&	0.0008&	-34 49 04.941&	0.013&	12.542&	0.029&	0.0042\\
10     355&	18 40 43.1574&	0.0008&	-34 45 44.037&	0.014&	12.691&	0.019&	0.0042\\
10     356&	18 40 49.8208&	0.0011&	-34 47 47.144&	0.012&	12.264&	0.018&	0.0036\\
10     357 - T&	18 40 58.0200&	0.0017&	-34 52 48.100&	0.011&	9.981&	0.022&	0.0048\\
10     358&	18 40 58.1509&	0.0018&	-34 53 28.222&	0.013&	12.587&	0.020&	0.0044\\
10     363&	18 41 16.3196&	0.0010&	-34 45 28.354&	0.011&	11.142&	0.018&	0.0052\\
10     364&	18 41 28.2108&	0.0012&	-34 54 14.767&	0.015&	12.874&	0.020&	0.0042\\
10     365&	18 41 34.2927&	0.0011&	-34 53 04.609&	0.013&	12.919&	0.027&	0.0047\\
10     366&	18 41 34.4927&	0.0015&	-34 47 53.102&	0.013&	11.986&	0.036&	0.0036\\
10     367&	18 41 35.1517&	0.0009&	-34 45 21.128&	0.012&	11.665&	0.019&	0.0031\\
10     369&	18 41 44.8033&	0.0050&	-34 48 29.378&	0.011&	11.782&	0.018&	0.0034\\
10     372&	18 42 16.3402&	0.0008&	-34 46 17.666&	0.015&	12.766&	0.017&	0.0041\\
10     373&	18 42 16.6841&	0.0007&	-34 53 32.551&	0.014&	11.994&	0.017&	0.0033\\
10     375 - T&	18 42 26.0900&	0.0007&	-34 44 28.799&	0.014&	10.413&	0.023&	0.0028\\
10     377&	18 42 28.6178&	0.0026&	-34 53 52.540&	0.015&	12.458&	0.020&	0.0044\\
10     378&	18 42 29.1555&	0.0006&	-34 49 49.044&	0.012&	12.813&	0.034&	0.0044\\
10     379&	18 42 36.1817&	0.0006&	-34 45 09.149&	0.013&	12.387&	0.024&	0.0040\\
10     380&	18 42 36.3324&	0.0014&	-34 48 23.589&	0.013&	11.748&	0.018&	0.0043\\
10     381&	18 42 38.2081&	0.0007&	-34 51 53.705&	0.015&	12.943&	0.042&	0.0047\\
10     382&	18 42 46.2861&	0.0036&	-34 48 39.102&	0.013&	12.944&	0.022&	0.0046\\
10     383&	18 42 49.7509&	0.0006&	-34 44 00.655&	0.015&	11.934&	0.030&	0.0036\\
10     384&	18 43 07.4668&	0.0008&	-34 51 38.161&	0.016&	12.641&	0.046&	0.0041\\
10     385&	18 43 11.7821&	0.0065&	-34 44 36.750&	0.009&	10.555&	0.022&	0.0055\\
10     386&	18 43 13.4761&	0.0010&	-34 47 19.003&	0.013&	12.818&	0.038&	0.0043\\
10     387&	18 43 15.6180&	0.0008&	-34 48 30.393&	0.013&	12.805&	0.024&	0.0043\\
10     388&	18 43 19.0595&	0.0013&	-34 44 33.964&	0.012&	12.416&	0.027&	0.0043\\
\hline    
\end{tabular}
\end{flushleft}
\end{table*}

\begin{table*}[tb]
\caption{Secondary catalog of references for the LA window (mean epoch 2450648, based on 33 nights
of observations)}
\label{tabrmm}
\begin{flushleft}
\begin{tabular}{@{}lcrcrccc@{}}
\hline
\rule{0pt}{1.2em}Name& $\alpha$ ($^{h}$ $^{m}$ $^{s}$)&$\sigma_{\alpha}$ ($^{s}$)&$\delta$ ($^{o}$ ' '')&$\sigma_{\delta} ('')$&$m_{Val}$&$\sigma_{m_{Val}}$&$\sigma_{w}$\\
\hline
11     509 &	16 54 26.0913 &	0.0008 &-52 41 30.159 &	0.009 &	11.949 &0.026 &	0.0032\\
11     510 &	16 54 31.0448 &	0.0028 &-52 44 40.393 &	0.053 &	12.655 &0.025 &	0.0037\\
11     511 &	16 54 33.0407 &	0.0009 &-52 48 29.344 &	0.009 &	12.966 &0.036 &	0.0058\\
11     515 &	16 54 38.4243 &	0.0015 &-52 44 40.337 &	0.009 &	12.138 &0.023 &	0.0038\\
11     516 &	16 54 44.7434 &	0.0035 &-52 50 06.575 &	0.010 &	12.547 &0.026 &	0.0037\\
11     518 &	16 54 54.7405 &	0.0008 &-52 46 09.499 &	0.009 &	12.416 &0.019 &	0.0055\\
11     519 &	16 54 56.4232 &	0.0005 &-52 47 29.986 &	0.008 &	12.088 &0.020 &	0.0026\\
11     522 &	16 55 09.4046 &	0.0006 &-52 43 33.422 &	0.010 &	13.028 &0.024 &	0.0038\\
11     523 &	16 55 11.1800 &	0.0007 &-52 42 54.474 &	0.012 &	12.796 &0.019 &	0.0034\\
11     525 &	16 55 14.7931 &	0.0006 &-52 40 52.436 &	0.012 &	12.579 &0.021 &	0.0033\\
11     527 &	16 55 23.7101 &	0.0007 &-52 45 22.443 &	0.010 &	12.835 &0.045 &	0.0042\\
11     529 &	16 55 32.8777 &	0.0011 &-52 40 23.440 &	0.013 &	12.675 &0.022 &	0.0040\\
11     531 &	16 55 36.3706 &	0.0043 &-52 40 38.219 &	0.011 &	12.939 &0.026 &	0.0040\\
11     532 &	16 55 40.0035 &	0.0007 &-52 49 08.697 &	0.009 &	10.943 &0.015 &	0.0042\\
11     536 &	16 55 44.8689 &	0.0010 &-52 47 52.646 &	0.009 &	11.269 &0.018 &	0.0040\\
11     537 &	16 55 44.8865 &	0.0007 &-52 46 55.745 &	0.010 &	11.804 &0.019 &	0.0028\\
11     538 &	16 55 56.7491 &	0.0006 &-52 45 47.344 &	0.009 &	11.895 &0.027 &	0.0029\\
11     542 - T &16 56 03.2300 &	0.0013 &-52 45 21.300 &	0.011 &	9.664 &	0.032 &	0.0068\\
11     545 &	16 56 13.7387 &	0.0050 &-52 44 12.251 &	0.010 &	12.590 &0.021 &	0.0036\\
11     549 &	16 56 17.7757 &	0.0008 &-52 45 59.225 &	0.011 &	12.374 &0.020 &	0.0030\\
11     550 &	16 56 20.3579 &	0.0015 &-52 42 08.422 &	0.011 &	11.869 &0.034 &	0.0039\\
11     551 &	16 56 28.1963 &	0.0025 &-52 40 54.324 &	0.011 &	11.702 &0.018 &	0.0033\\
11     552 &	16 56 36.3083 &	0.0007 &-52 49 31.285 &	0.012 &	11.981 &0.021 &	0.0038\\
11     560 &	16 57 05.7584 &	0.0020 &-52 43 41.347 &	0.012 &	12.488 &0.021 &	0.0033\\
11     561 &	16 57 06.4307 &	0.0010 &-52 48 01.479 &	0.013 &	12.939 &0.028 &	0.0040\\
11     562 &	16 57 06.5173 &	0.0010 &-52 48 36.013 &	0.010 &	12.762 &0.026 &	0.0039\\
11     563 &	16 57 10.8121 &	0.0007 &-52 42 40.679 &	0.010 &	12.002 &0.024 &	0.0036\\
11     564 - T &16 57 11.2400 &	0.0052 &-52 48 11.600 &	0.010 &	10.557 &0.030 &	0.0046\\
11     565 & 	16 57 18.1550 &	0.0194 &-52 43 44.770 &	0.011 &	12.543 &0.026 &	0.0039\\
11     567 &	16 57 21.4232 &	0.0008 &-52 41 39.526 &	0.009 &	12.533 &0.036 &	0.0032\\
11     568 - T &16 57 23.8500 &	0.0045 &-52 42 08.200 &	0.010 &	10.496 &0.031 &	0.0022\\
11     569 &	16 57 26.2187 &	0.0011 &-52 50 02.611 &	0.013 &	12.947 &0.027 &	0.0042\\
11     570 &	16 57 31.6293 &	0.0021 &-52 49 26.490 &	0.011 &	12.039 &0.018 &	0.0029\\
11     571 &	16 57 33.1571 &	0.0013 &-52 45 04.694 &	0.013 &	11.867 &0.022 &	0.0031\\
11     573 - T &16 57 47.1400 &	0.0011 &-52 42 23.200 &	0.010 &	11.164 &0.017 &	0.0029\\
11     574 - T &16 57 50.7999 &	0.0006 &-52 42 22.700 &	0.010 &	11.279 &0.016 &	0.0034\\
11     575 &	16 57 51.1421 &	0.0007 &-52 45 22.814 &	0.009 &	11.486 &0.019 &	0.0028\\
11     576 &	16 57 52.1364 &	0.0008 &-52 40 42.571 &	0.011 &	12.832 &0.037 &	0.0039\\
11     579 &	16 58 09.4619 &	0.0021 &-52 41 46.512 &	0.009 &	12.746 &0.025 &	0.0037\\
11     581 - T &16 58 11.4399 &	0.0050 &-52 48 36.500 &	0.008 &	9.905 &	0.033 &	0.0052\\
11     585 - T &16 58 16.2300 &	0.0009 &-52 47 41.499 &	0.010 &	10.363 &0.033 &	0.0034\\
11     590 &	16 58 37.4512 &	0.0006 &-52 42 48.134 &	0.010 &	12.804 &0.043 &	0.0051\\
11     591 &	16 58 41.5723 &	0.0005 &-52 43 30.962 &	0.010 &	12.908 &0.048 &	0.0043\\	
11     594 &	16 58 53.3302 &	0.0004 &-52 41 54.019 &	0.009 &	12.741 &0.065 &	0.0032\\
11     595 &	16 58 55.7209 &	0.0013 &-52 42 39.827 &	0.007 &	12.806 &0.050 &	0.0033\\
11     596 &	16 58 55.8769 &	0.0008 &-52 42 21.094 &	0.009 &	12.467 &0.058 &	0.0032\\
11     598 &	16 59 02.7172 &	0.0007 &-52 48 42.319 &	0.010 &	12.145 &0.059 &	0.0038\\
11     602 &	16 59 05.8484 &	0.0003 &-52 41 04.356 &	0.006 &	12.453 &0.056 &	0.0032\\
11     603 &	16 59 06.1039 &	0.0005 &-52 39 55.771 &	0.008 &	12.819 &0.075 &	0.0039\\
11     606 &	16 59 09.8720 &	0.0040 &-52 41 49.858 &	0.007 &	12.748 &0.078 &	0.0034\\
11     608 &	16 59 12.1369 &	0.0034 &-52 41 06.077 &	0.009 &	12.532 &0.065 &	0.0034\\
11     610 &	16 59 15.9486 &	0.0008 &-52 42 26.126 &	0.008 &	12.765 &0.073 &	0.0036\\
11     615 &	16 59 24.6255 &	0.0045 &-52 44 18.518 &	0.008 &	12.549 &0.076 &	0.0032\\
11     618 - T &16 59 31.2399 &	0.0014 &-52 46 41.199 &	0.010 &	10.562 &0.080 &	0.0048\\
11     619 &	16 59 32.7690 &	0.0007 &-52 44 05.387 &	0.012 &	12.715 &0.098 &	0.0046\\
\hline    
\end{tabular}
\end{flushleft}
\end{table*}

\begin{table*}[tb]
\caption{Secondary catalog of references for the LB window (mean epoch 2450657, based on 27 nights
of observations)}
\label{tabrmm}
\begin{flushleft}
\begin{tabular}{@{}lcrcrccc@{}}
\hline
\rule{0pt}{1.2em}Name& $\alpha$ ($^{h}$ $^{m}$ $^{s}$)&$\sigma_{\alpha}$ ($^{s}$)&$\delta$ ($^{o}$ ' '')&$\sigma_{\delta} ('')$&$m_{Val}$&$\sigma_{m_{Val}}$&$\sigma_{w}$\\
\hline
12     394 &	17 15 15.5119&	0.0020&	-48 08 58.223&	0.010&	12.995&	0.030&	0.0045\\
12     396 &	17 15 30.5699&	0.0015&	-48 10 02.838&	0.012&	12.189&	0.050&	0.0040\\
12     400 &	17 15 38.4216&	0.0086&	-48 10 15.742&	0.009&	12.728&	0.035&	0.0042\\
12     401 &	17 15 45.6920&	0.0024&	-48 06 22.748&	0.012&	12.393&	0.019&	0.0037\\
12     403 &	17 16 00.5613&	0.0006&	-48 05 12.444&	0.012&	12.921&	0.032&	0.0043\\
12     404 &	17 16 22.0083&	0.0006&	-48 03 02.528&	0.013&	12.132&	0.032&	0.0039\\
12     406 &	17 16 33.0385&	0.0011&	-48 02 07.631&	0.011&	12.938&	0.074&	0.0042\\
12     407 &	17 16 33.1861&	0.0010&	-48 05 12.158&	0.014&	12.900&	0.035&	0.0050\\
12     408 &	17 16 33.7440&	0.0009&	-48 07 36.966&	0.011&	12.267&	0.037&	0.0031\\
12     410 &	17 16 45.2348&	0.0006&	-48 04 54.929&	0.009&	12.124&	0.027&	0.0032\\
12     415 &	17 17 10.5838&	0.0035&	-48 07 18.979&	0.013&	12.758&	0.042&	0.0044\\
12     416 &	17 17 11.8075&	0.0008&	-48 08 23.142&	0.012&	11.604&	0.021&	0.0032\\
12     420 &	17 17 33.5453&	0.0010&	-48 08 41.362&	0.014&	12.792&	0.044&	0.0054\\
12     429 - T &17 17 59.1483&	0.0023&	-48 03 20.000&	0.012&	11.069&	0.041&	0.0043\\
12     430 &	17 18 01.2958&	0.0043&	-48 06 23.158&	0.015&	12.757&	0.035&	0.0044\\
12     432 &	17 18 12.3685&	0.0009&	-48 04 14.348&	0.013&	12.977&	0.030&	0.0041\\
12     434 &	17 18 19.0746&	0.0072&	-48 04 43.921&	0.011&	13.006&	0.046&	0.0040\\
12     435 & 	17 18 22.2706&	0.0007&	-48 05 32.009&	0.012&	11.898&	0.025&	0.0034\\
12     438 &	17 18 31.3619&	0.0015&	-48 04 15.137&	0.011&	10.769&	0.027&	0.0051\\
12     441 &	17 18 39.8959&	0.0007&	-48 03 33.917&	0.012&	11.780&	0.025&	0.0029\\
12     447 &	17 19 08.4432&	0.0007&	-48 09 41.037&	0.009&	11.630&	0.018&	0.0029\\
12     450 &	17 19 17.1911&	0.0009&	-48 05 48.371&	0.012&	12.543&	0.023&	0.0035\\
12     451 &	17 19 20.6107&	0.0008&	-48 10 02.470&	0.010&	12.857&	0.030&	0.0041\\
12     452 &	17 19 24.9759&	0.0027&	-48 09 13.490&	0.010&	11.307&	0.020&	0.0028\\
12     454 & 	17 19 29.4815&	0.0009&	-48 09 41.243&	0.010&	12.889&	0.026&	0.0040\\
12     455 &	17 19 32.6798&	0.0011&	-48 02 58.828&	0.009&	12.751&	0.035&	0.0042\\
12     456 - T &17 19 41.5898&	0.0007&	-48 05 59.500&	0.010&	10.348&	0.037&	0.0055\\
12     458 - T &17 19 44.5391&	0.0080&	-48 08 33.500&	0.008&	10.928&	0.025&	0.0035\\
12     459 - T &17 19 46.9805&	0.0006&	-48 01 30.297&	0.008&	11.178&	0.061&	0.0034\\
12     460 &	17 19 48.0964&	0.0008&	-48 10 26.580&	0.013&	12.536&	0.025&	0.0039\\
12     461 &	17 19 57.6453&	0.0006&	-48 05 56.951&	0.009&	11.988&	0.019&	0.0031\\
12     462 &	17 20 00.3225&	0.0045&	-48 02 39.403&	0.011&	12.780&	0.048&	0.0035\\
12     465 &	17 20 05.3706&	0.0011&	-48 06 56.904&	0.012&	11.893&	0.036&	0.0029\\
12     466 &	17 20 11.6808&	0.0015&	-48 05 01.553&	0.012&	12.698&	0.024&	0.0037\\
12     467 &	17 20 14.0748&	0.0006&	-48 05 41.055&	0.011&	11.762&	0.022&	0.0032\\
12     468 &	17 20 24.7949&	0.0008&	-48 02 54.339&	0.011&	11.608&	0.037&	0.0048\\
\hline
\end{tabular}
\end{flushleft}
\end{table*}

\begin{table*}[tb]
\caption{Secondary catalog of references for the LC window (mean epoch 2450653, based on 37 nights
of observations)}
\label{tabrmm}
\begin{flushleft}
\begin{tabular}{@{}lcrcrccc@{}}
\hline
\rule{0pt}{1.2em}Name& $\alpha$ ($^{h}$ $^{m}$ $^{s}$)&$\sigma_{\alpha}$ ($^{s}$)&$\delta$ ($^{o}$ ' '')&$\sigma_{\delta} ('')$&$m_{Val}$&$\sigma_{m_{Val}}$&$\sigma_{w}$\\
\hline
13     251&	17 27 01.3073&	0.0006&	-45 19 29.615&	0.010&	12.021&	0.024&	0.0029\\
13     252 - T&	17 27 09.6797&	0.0005&	-45 24 31.594&	0.008&	10.014&	0.019&	0.0052\\
13     256&	17 27 24.3658&	0.0081&	-45 21 58.167&	0.011&	12.689&	0.035&	0.0034\\
13     257&	17 27 28.3251&	0.0011&	-45 20 30.232&	0.012&	12.340&	0.028&	0.0035\\
13     258&	17 27 30.6504&	0.0006&	-45 20 50.955&	0.011&	11.312&	0.030&	0.0027\\
13     259&	17 27 31.5059&	0.0007&	-45 21 24.497&	0.010&	11.939&	0.023&	0.0028\\
13     260&	17 27 32.3165&	0.0011&	-45 23 58.535&	0.013&	12.815&	0.028&	0.0034\\
13     262&	17 27 36.3602&	0.0006&	-45 15 21.609&	0.011&	12.484&	0.031&	0.0036\\
13     263&	17 27 38.0735&	0.0009&	-45 20 58.426&	0.010&	10.706&	0.032&	0.0046\\
13     264&	17 27 38.9241&	0.0008&	-45 24 40.248&	0.010&	11.015&	0.022&	0.0057\\
13     266&	17 27 40.6048&	0.0005&	-45 21 25.302&	0.011&	12.007&	0.027&	0.0030\\
13     269&	17 27 45.3995&	0.0007&	-45 22 42.543&	0.012&	12.382&	0.034&	0.0032\\
13     270&	17 27 52.0152&	0.0010&	-45 15 02.000&	0.010&	13.059&	0.046&	0.0041\\
13     272&	17 28 00.2731&	0.0024&	-45 17 42.932&	0.012&	12.973&	0.026&	0.0035\\
13     273&	17 28 00.5315&	0.0014&	-45 17 06.099&	0.011&	11.688&	0.019&	0.0028\\
13     274&	17 28 00.6479&	0.0021&	-45 18 13.571&	0.021&	12.783&	0.034&	0.0040\\
13     275&	17 28 00.8287&	0.0008&	-45 19 18.093&	0.010&	12.251&	0.027&	0.0030\\
13     277&	17 28 03.1251&	0.0007&	-45 20 17.068&	0.010&	12.019&	0.019&	0.0030\\
13     278&	17 28 08.6793&	0.0055&	-45 20 46.404&	0.010&	12.685&	0.029&	0.0035\\
13     280&	17 28 12.9593&	0.0056&	-45 17 24.537&	0.013&	12.963&	0.033&	0.0036\\
13     281&	17 28 13.3725&	0.0007&	-45 22 52.205&	0.009&	12.622&	0.026&	0.0033\\
13     282&	17 28 20.0064&	0.0011&	-45 22 07.510&	0.011&	12.233&	0.024&	0.0034\\
13     283&	17 28 23.3062&	0.0008&	-45 17 31.508&	0.010&	12.074&	0.029&	0.0038\\
13     284&	17 28 23.6751&	0.0007&	-45 16 42.806&	0.012&	10.607&	0.030&	0.0017\\
13     287&	17 28 28.7178&	0.0007&	-45 21 34.714&	0.013&	12.816&	0.030&	0.0034\\
13     289 - T&	17 28 33.8594&	0.0009&	-45 18 44.703&	0.010&	9.382&	0.031&	0.0041\\
13     290&	17 28 53.1246&	0.0007&	-45 20 48.400&	0.013&	12.082&	0.021&	0.0037\\
13     293&	17 29 06.8373&	0.0021&	-45 17 43.609&	0.011&	12.699&	0.025&	0.0036\\
13     294&	17 29 09.8707&	0.0007&	-45 17 03.657&	0.012&	12.848&	0.047&	0.0034\\
13     295&	17 29 12.0071&	0.0025&	-45 22 56.855&	0.012&	12.426&	0.030&	0.0039\\
13     297&	17 29 12.5162&	0.0007&	-45 24 02.839&	0.012&	12.034&	0.021&	0.0026\\
13     301 - T&	17 29 22.6094&	0.0011&	-45 16 19.500&	0.011&	10.664&	0.026&	0.0023\\
13     303&	17 29 23.5523&	0.0016&	-45 20 49.236&	0.009&	11.738&	0.022&	0.0030\\
13     305&	17 29 33.8267&	0.0006&	-45 15 02.578&	0.010&	11.732&	0.030&	0.0030\\
13     306&	17 29 38.0649&	0.0010&	-45 15 47.695&	0.011&	11.057&	0.020&	0.0045\\
13     307&	17 29 41.9061&	0.0019&	-45 23 40.706&	0.010&	12.690&	0.024&	0.0033\\
13     308 - T&	17 29 42.6211&	0.0015&	-45 23 06.297&	0.012&	9.966&	0.036&	0.0048\\
13     311&	17 29 51.5479&	0.0027&	-45 20 27.878&	0.010&	12.147&	0.025&	0.0032\\
13     312&	17 29 52.3801&	0.0006&	-45 22 14.247&	0.011&	12.289&	0.026&	0.0030\\
13     313&	17 29 54.6947&	0.0007&	-45 22 10.928&	0.009&	11.300&	0.023&	0.0034\\
13     314&	17 29 58.2236&	0.0009&	-45 17 40.652&	0.008&	12.103&	0.028&	0.0029\\
13     315&	17 29 59.8391&	0.0007&	-45 23 00.669&	0.010&	12.916&	0.033&	0.0033\\
13     316&	17 30 15.5759&	0.0007&	-45 16 44.879&	0.010&	11.518&	0.020&	0.0043\\
13     317&	17 30 19.0175&	0.0006&	-45 18 12.931&	0.010&	12.176&	0.027&	0.0038\\
13     320&	17 30 23.1773&	0.0006&	-45 15 59.777&	0.011&	12.941&	0.034&	0.0039\\
13     321&	17 30 24.8896&	0.0006&	-45 23 31.782&	0.010&	11.616&	0.020&	0.0025\\
13     322 - T&	17 30 25.6094&	0.0010&	-45 17 45.906&	0.009&	10.909&	0.032&	0.0024\\
13     323&	17 30 30.2458&	0.0072&	-45 18 17.027&	0.012&	12.942&	0.036&	0.0038\\
13     324&	17 30 34.0148&	0.0005&	-45 20 51.537&	0.013&	12.462&	0.023&	0.0029\\
13     325&	17 30 36.5155&	0.0011&	-45 21 49.160&	0.011&	11.507&	0.032&	0.0029\\
13     326&	17 30 38.0049&	0.0022&	-45 25 01.025&	0.013&	12.105&	0.083&	0.0031\\
13     327&	17 30 39.7271&	0.0017&	-45 20 06.012&	0.011&	11.759&	0.017&	0.0028\\
13     328 - T&	17 30 44.1623&	0.0006&	-45 21 15.434&	0.009&	10.397&	0.018&	0.0047\\
13     329&	17 30 45.1443&	0.0044&	-45 16 15.817&	0.009&	12.834&	0.031&	0.0043\\
13     330&	17 30 45.7255&	0.0008&	-45 21 43.171&	0.009&	11.362&	0.019&	0.0033\\
13     331 - T&	17 30 46.5313&	0.0016&	-45 20 14.797&	0.007&	10.098&	0.030&	0.0062\\
13     336&	17 31 14.8788&	0.0005&	-45 17 47.968&	0.011&	12.863&	0.026&	0.0034\\
13     337&	17 31 19.1029&	0.0006&	-45 22 05.254&	0.010&	10.701&	0.025&	0.0029\\
13     338&	17 31 23.3392&	0.0009&	-45 14 59.610&	0.011&	12.440&	0.055&	0.0031\\
13     340 - T&	17 31 27.8789&	0.0007&	-45 25 11.500&	0.010&	10.998&	0.029&	0.0034\\
13     341&	17 31 32.6591&	0.0007&	-45 16 39.370&	0.011&	12.275&	0.034&	0.0044\\
\hline
\end{tabular}
\end{flushleft}
\end{table*}

\begin{table*}[tb]
\caption{Secondary catalog of references for the LD window (mean epoch 2450654, based on 26 nights
of observations)}
\label{tabrmm}
\begin{flushleft}
\begin{tabular}{@{}lcrcrccc@{}}
\hline
\rule{0pt}{1.2em}Name& $\alpha$ ($^{h}$ $^{m}$ $^{s}$)&$\sigma_{\alpha}$ ($^{s}$)&$\delta$ ($^{o}$ ' '')&$\sigma_{\delta} ('')$&$m_{Val}$&$\sigma_{m_{Val}}$&$\sigma_{w}$\\
\hline
14     413&	17 42 17.9627&	0.0007&	-40 37 29.144&	0.011&	12.903&	0.030&	0.0038\\
14     415&	17 42 24.1889&	0.0007&	-40 46 19.925&	0.011&	12.430&	0.024&	0.0035\\
14     416&	17 42 27.3763&	0.0008&	-40 41 29.935&	0.010&	12.403&	0.030&	0.0031\\
14     417&	17 42 27.3837&	0.0014&	-40 43 03.722&	0.010&	11.969&	0.025&	0.0027\\
14     418&	17 42 31.4609&	0.0004&	-40 45 08.852&	0.011&	12.098&	0.022&	0.0034\\
14     419 - T&	17 42 36.0195&	0.0044&	-40 38 08.297&	0.009&	10.288&	0.015&	0.0052\\
14     421&	17 42 41.1771&	0.0005&	-40 45 44.046&	0.013&	12.805&	0.031&	0.0040\\
14     423&	17 42 51.6847&	0.0005&	-40 40 30.119&	0.011&	12.842&	0.025&	0.0036\\
14     424&	17 42 54.0227&	0.0007&	-40 36 59.845&	0.009&	12.432&	0.034&	0.0038\\
14     426&	17 43 00.2661&	0.0011&	-40 39 14.281&	0.010&	11.118&	0.028&	0.0021\\
14     427&	17 43 00.7099&	0.0006&	-40 37 04.739&	0.010&	12.419&	0.021&	0.0039\\
14     428&	17 43 01.8737&	0.0007&	-40 45 53.871&	0.009&	11.277&	0.027&	0.0025\\
14     429&	17 43 04.2366&	0.0005&	-40 43 05.006&	0.009&	10.718&	0.019&	0.0035\\
14     430&	17 43 09.4483&	0.0145&	-40 44 14.928&	0.010&	12.175&	0.020&	0.0034\\
14     432&	17 43 11.9331&	0.0006&	-40 45 02.374&	0.008&	11.213&	0.032&	0.0021\\
14     433&	17 43 12.4672&	0.0032&	-40 39 13.867&	0.007&	11.198&	0.021&	0.0039\\
14     434&	17 43 16.0372&	0.0006&	-40 46 08.109&	0.008&	11.897&	0.022&	0.0027\\
14     435 - T&	17 43 19.0117&	0.0006&	-40 41 04.703&	0.010&	10.145&	0.062&	0.0030\\
14     437&	17 43 21.6065&	0.0019&	-40 37 50.522&	0.008&	11.667&	0.021&	0.0028\\
14     440&	17 43 39.1045&	0.0007&	-40 45 24.523&	0.011&	12.626&	0.028&	0.0036\\
14     441&	17 43 40.7168&	0.0007&	-40 38 25.844&	0.011&	12.516&	0.027&	0.0036\\
14     443&	17 43 47.5508&	0.0009&	-40 45 57.258&	0.011&	12.437&	0.041&	0.0032\\
14     447&	17 43 54.7157&	0.0008&	-40 41 57.990&	0.014&	12.651&	0.033&	0.0036\\
14     448&	17 44 01.1751&	0.0006&	-40 38 56.443&	0.012&	12.918&	0.025&	0.0039\\
14     450&	17 44 04.8167&	0.0005&	-40 46 24.023&	0.011&	10.829&	0.016&	0.0065\\
14     451&	17 44 08.5161&	0.0009&	-40 40 02.725&	0.014&	11.439&	0.016&	0.0037\\
14     452&	17 44 08.8959&	0.0008&	-40 40 43.494&	0.014&	12.622&	0.025&	0.0034\\
14     453&	17 44 12.7080&	0.0006&	-40 44 28.310&	0.013&	12.006&	0.021&	0.0029\\
14     454&	17 44 20.9446&	0.0006&	-40 37 05.250&	0.012&	12.703&	0.022&	0.0034\\
14     459&	17 44 53.6561&	0.0005&	-40 45 56.229&	0.016&	12.589&	0.025&	0.0035\\
14     460 - T&	17 44 58.9492&	0.0017&	-40 41 37.094&	0.010&	9.926&	0.030&	0.0025\\
14     461&	17 45 00.6384&	0.0009&	-40 43 04.067&	0.012&	12.256&	0.025&	0.0034\\
14     467&	17 45 29.7191&	0.0023&	-40 46 38.518&	0.017&	12.794&	0.039&	0.0055\\
14     468&	17 45 31.0759&	0.0006&	-40 44 09.482&	0.010&	12.509&	0.023&	0.0037\\
14     469&	17 45 39.1752&	0.0066&	-40 41 30.436&	0.010&	12.204&	0.035&	0.0055\\
14     470&	17 45 43.4305&	0.0007&	-40 40 10.465&	0.008&	12.456&	0.024&	0.0037\\
14     471&	17 45 53.3889&	0.0011&	-40 40 38.228&	0.010&	11.195&	0.022&	0.0027\\
14     472&	17 45 54.6981&	0.0020&	-40 44 24.654&	0.010&	11.741&	0.053&	0.0042\\
14     475&	17 45 56.9327&	0.0005&	-40 43 39.648&	0.009&	12.428&	0.028&	0.0035\\
14     476&	17 45 57.0257&	0.0007&	-40 43 18.080&	0.009&	12.549&	0.043&	0.0048\\
14     477&	17 45 57.2522&	0.0005&	-40 41 41.454&	0.010&	12.656&	0.035&	0.0036\\
14     478&	17 45 59.0025&	0.0045&	-40 39 13.542&	0.011&	12.908&	0.024&	0.0038\\
14     483&	17 46 07.9966&	0.0010&	-40 46 19.983&	0.011&	12.963&	0.020&	0.0044\\
14     484&	17 46 12.7326&	0.0014&	-40 40 12.292&	0.013&	12.619&	0.031&	0.0035\\
14     488&	17 46 22.9859&	0.0005&	-40 38 53.958&	0.013&	12.305&	0.019&	0.0033\\
14     489 - T&	17 46 23.1211&	0.0018&	-40 38 02.406&	0.011&	10.096&	0.013&	0.0044\\
14     490&	17 46 29.4764&	0.0007&	-40 40 29.512&	0.014&	12.618&	0.032&	0.0037\\
\hline
\end{tabular}
\end{flushleft}
\end{table*}

\begin{table*}[tb]
\caption{Secondary catalog of references for the LI window (mean epoch 2450637, based on 34 nights
of observations)}
\label{tabrmm}
\begin{flushleft}
\begin{tabular}{@{}lcrcrccc@{}}
\hline
\rule{0pt}{1.2em}Name& $\alpha$ ($^{h}$ $^{m}$ $^{s}$)&$\sigma_{\alpha}$ ($^{s}$)&$\delta$ ($^{o}$ ' '')&$\sigma_{\delta} ('')$&$m_{Val}$&$\sigma_{m_{Val}}$&$\sigma_{w}$\\
\hline
19     474 - T&	18  03 13.7109&	0.0006&	-33 50 13.703&	0.008&	10.115&	0.024&	0.0048\\
19     477&	18  03 20.8416&	0.0004&	-33 46 46.813&	0.008&	11.914&	0.057&	0.0028\\
19     478&	18  03 24.6165&	0.0006&	-33 50 14.775&	0.010&	12.591&	0.046&	0.0043\\
19     479&	18  03 29.1408&	0.0008&	-33 50 13.527&	0.005&	10.925&	0.019&	0.0029\\
19     480&	18  03 35.9087&	0.0022&	-33 50 28.775&	0.009&	12.744&	0.022&	0.0037\\
19     481&	18  03 40.4583&	0.0025&	-33 51 21.461&	0.013&	12.985&	0.031&	0.0046\\
19     482&	18  03 40.7783&	0.0006&	-33 49 44.052&	0.008&	12.442&	0.035&	0.0045\\
19     483&	18  03 58.2167&	0.0007&	-33 44 17.984&	0.010&	11.940&	0.022&	0.0033\\
19     487&	18  04 19.7327&	0.0006&	-33 48 13.171&	0.013&	12.085&	0.024&	0.0034\\
19     488&	18  04 29.6119&	0.0024&	-33 45 20.736&	0.012&	12.272&	0.023&	0.0033\\
19     490&	18  04 47.2949&	0.0007&	-33 52 05.961&	0.012&	12.046&	0.024&	0.0030\\
19     491&	18  04 47.4174&	0.0008&	-33 48 51.479&	0.011&	12.506&	0.031&	0.0034\\
19     493&	18  04 53.1317&	0.0012&	-33 49 54.087&	0.015&	12.483&	0.027&	0.0044\\
19     496&	18  05 03.5905&	0.0007&	-33 47 45.170&	0.011&	12.560&	0.085&	0.0046\\
19     498&	18  05 11.0333&	0.0006&	-33 47 20.107&	0.008&	12.934&	0.029&	0.0041\\
19     500&	18  05 15.2713&	0.0004&	-33 45 38.584&	0.008&	12.198&	0.022&	0.0031\\
19     502&	18  05 28.0544&	0.0007&	-33 51 16.629&	0.015&	12.942&	0.035&	0.0046\\
19     504&	18  05 36.6298&	0.0008&	-33 52 34.616&	0.013&	11.973&	0.026&	0.0038\\
19     505&	18  05 38.4969&	0.0008&	-33 45 32.015&	0.012&	11.366&	0.022&	0.0030\\
19     506&	18  05 40.7910&	0.0014&	-33 52 41.592&	0.012&	12.934&	0.023&	0.0042\\
\hline
\end{tabular}
\end{flushleft}
\end{table*}

\begin{table*}[tb]
\caption{Secondary catalog of references for the LR window (mean epoch 2450617, based on 24 nights
of observations)}
\label{tabrmm}
\begin{flushleft}
\begin{tabular}{@{}lcrcrccc@{}}
\hline
\rule{0pt}{1.2em}Name& $\alpha$ ($^{h}$ $^{m}$ $^{s}$)&$\sigma_{\alpha}$ ($^{s}$)&$\delta$ ($^{o}$ ' '')&$\sigma_{\delta} ('')$&$m_{Val}$&$\sigma_{m_{Val}}$&$\sigma_{w}$\\
\hline
27     456& 	18 22 17.8644&	0.0018&	-26 08 27.730&	0.015&	13.011&	0.080&	0.0059\\
27     457& 	18 22 18.2137&	0.0112&	-26 09 51.334&	0.012&	11.867&	0.072&	0.0051\\
27     460& 	18 22 30.4714&	0.0020&	-26 10 09.285&	0.015&	12.786&	0.020&	0.0046\\
27     461& 	18 22 32.4742&	0.0010&	-26 05 05.630&	0.016&	11.659&	0.024&	0.0050\\
27     463& 	18 22 41.7036&	0.0009&	-26 12 05.535&	0.014&	12.262&	0.034&	0.0039\\
27     465& 	18 22 53.4489&	0.0017&	-26 14 23.732&	0.013&	12.033&	0.022&	0.0038\\
27     467& 	18 22 56.5578&	0.0011&	-26 11 10.427&	0.013&	12.326&	0.018&	0.0042\\
27     470& 	18 23 04.6275&	0.0017&	-26 05 15.086&	0.014&	12.814&	0.039&	0.0054\\
27     471& 	18 23 07.2110&	0.0007&	-26 04 32.941&	0.019&	12.754&	0.030&	0.0054\\
27     472& 	18 23 08.9807&	0.0098&	-26 06 22.882&	0.020&	12.140&	0.036&	0.0062\\
27     474 - T& 18 23 27.2969&	0.0009&	-26 05 06.008&	0.017&	10.418&	0.030&	0.0027\\
27     477& 	18 23 46.9881&	0.0012&	-26 12 46.417&	0.015&	12.852&	0.035&	0.0048\\
27     478 - T& 18 24 04.0625&	0.0009&	-26 07 45.273&	0.018&	10.234&	0.022&	0.0047\\
27     480& 	18 24 10.8828&	0.0014&	-26 11 17.291&	0.018&	11.630&	0.021&	0.0046\\
27     483& 	18 24 25.5686&	0.0026&	-26 09 04.589&	0.012&	12.466&	0.019&	0.0039\\
27     486& 	18 24 34.0611&	0.0027&	-26 14 46.071&	0.010&	12.479&	0.030&	0.0052\\
27     487& 	18 24 38.8575&	0.0054&	-26 09 22.995&	0.014&	12.427&	0.019&	0.0045\\
27     490& 	18 24 41.7854&	0.0010&	-26 15 18.622&	0.013&	12.316&	0.032&	0.0053\\
27     491 - T& 18 24 41.8203&	0.0026&	-26 13 43.977&	0.015&	10.551&	0.020&	0.0055\\
27     492& 	18 24 43.7855&	0.0011&	-26 05 13.716&	0.011&	11.258&	0.025&	0.0054\\
27     493& 	18 24 45.4045&	0.0011&	-26 06 31.642&	0.011&	11.989&	0.023&	0.0042\\
27     494& 	18 24 50.6043&	0.0007&	-26 10 48.388&	0.013&	12.039&	0.021&	0.0034\\
27     495& 	18 24 52.2755&	0.0016&	-26 06 57.095&	0.009&	11.032&	0.036&	0.0031\\
27     497& 	18 24 57.3874&	0.0018&	-26 07 27.838&	0.011&	12.666&	0.036&	0.0050\\
27     499& 	18 24 59.2453&	0.0012&	-26 10 46.677&	0.011&	12.858&	0.034&	0.0054\\
27     500& 	18 24 59.6887&	0.0033&	-26 09 01.370&	0.009&	12.629&	0.044&	0.0058\\
27     502& 	18 25 04.5005&	0.0006&	-26 07 42.758&	0.011&	12.550&	0.023&	0.0043\\
27     503& 	18 25 05.5788&	0.0064&	-26 10 47.336&	0.012&	12.745&	0.027&	0.0050\\
27     504& 	18 25 11.6961&	0.0012&	-26 15 22.898&	0.013&	12.516&	0.024&	0.0048\\
27     505& 	18 25 16.5980&	0.0122&	-26 05 05.068&	0.016&	12.759&	0.043&	0.0052\\
27     507& 	18 25 21.4535&	0.0007&	-26 13 21.299&	0.013&	11.072&	0.027&	0.0043\\
27     508& 	18 25 29.1589&	0.0025&	-26 09 45.384&	0.016&	12.915&	0.026&	0.0050\\
27     509& 	18 25 29.3665&	0.0034&	-26 14 02.588&	0.017&	12.893&	0.024&	0.0047\\
27     510& 	18 25 29.9184&	0.0008&	-26 04 32.552&	0.017&	12.850&	0.037&	0.0047\\
27     511& 	18 25 39.1315&	0.0009&	-26 09 34.263&	0.019&	12.996&	0.036&	0.0055\\
27     512& 	18 25 41.0147&	0.0009&	-26 05 19.297&	0.015&	12.828&	0.040&	0.0055\\
27     513& 	18 25 41.6689&	0.0011&	-26 05 59.622&	0.018&	11.973&	0.019&	0.0043\\
\hline 
\end{tabular}
\end{flushleft}
\end{table*}

\begin{table*}[tb]
\caption{Secondary catalog of references for the LT window (mean epoch 2450610, based on 25 nights
of observations)}
\label{tabrmm}
\begin{flushleft}
\begin{tabular}{@{}lcrcrccc@{}}
\hline
\rule{0pt}{1.2em}Name& $\alpha$ ($^{h}$ $^{m}$ $^{s}$)&$\sigma_{\alpha}$ ($^{s}$)&$\delta$ ($^{o}$ ' '')&$\sigma_{\delta} ('')$&$m_{Val}$&$\sigma_{m_{Val}}$&$\sigma_{w}$\\
\hline
29     373&	18 32 25.8246&	0.0009&	-21 22 51.797&	0.013&	12.711&	0.023&	0.0050\\
29     375&	18 32 31.6099&	0.0008&	-21 25 41.713&	0.010&	11.331&	0.042&	0.0053\\
29     376&	18 32 35.2759&	0.0020&	-21 19 52.293&	0.011&	11.650&	0.032&	0.0036\\
29     377&	18 32 39.9934&	0.0013&	-21 26 33.812&	0.013&	12.622&	0.020&	0.0043\\
29     378&	18 32 42.5765&	0.0012&	-21 22 11.414&	0.015&	12.721&	0.026&	0.0047\\
29     379&	18 32 42.8892&	0.0176&	-21 25 49.864&	0.013&	12.987&	0.038&	0.0057\\
29     381&	18 32 46.9974&	0.0019&	-21 23 23.066&	0.012&	12.593&	0.025&	0.0046\\
29     382&	18 32 50.1448&	0.0015&	-21 25 58.786&	0.012&	11.310&	0.044&	0.0058\\
29     383&	18 32 57.5714&	0.0052&	-21 26 48.689&	0.010&	11.641&	0.027&	0.0057\\
29     384&	18 32 58.7671&	0.0021&	-21 27 07.367&	0.016&	12.753&	0.035&	0.0053\\
29     385&	18 33 11.7104&	0.0008&	-21 25 04.929&	0.010&	11.380&	0.024&	0.0047\\
29     386&	18 33 19.4227&	0.0013&	-21 24 53.215&	0.012&	12.143&	0.026&	0.0039\\
29     388&	18 33 20.3519&	0.0011&	-21 24 24.873&	0.012&	11.943&	0.020&	0.0040\\
29     389&	18 33 28.3465&	0.0017&	-21 27 42.062&	0.018&	12.662&	0.021&	0.0053\\
29     390&	18 33 38.1953&	0.0011&	-21 20 05.992&	0.014&	12.339&	0.024&	0.0055\\
29     391&	18 33 38.2212&	0.0018&	-21 26 05.415&	0.014&	11.995&	0.023&	0.0042\\
29     392&	18 33 38.4465&	0.0030&	-21 24 37.846&	0.014&	12.872&	0.020&	0.0058\\
29     394&	18 33 43.2223&	0.0012&	-21 18 49.141&	0.014&	12.371&	0.032&	0.0046\\
29     399&	18 33 52.6675&	0.0010&	-21 25 00.708&	0.017&	11.538&	0.029&	0.0042\\
29     400 - T&	18 34 01.5703&	0.0009&	-21 18 05.539&	0.013&	10.029&	0.029&	0.0082\\
29     401 - T&	18 34 08.3901&	0.0010&	-21 26 41.328&	0.016&	10.511&	0.023&	0.0059\\
29     402&	18 34 14.4232&	0.0009&	-21 26 30.911&	0.018&	12.795&	0.043&	0.0048\\
29     406&	18 34 27.4651&	0.0023&	-21 22 49.286&	0.014&	12.102&	0.021&	0.0039\\
29     411&	18 34 48.5393&	0.0018&	-21 27 45.149&	0.009&	11.713&	0.020&	0.0035\\
29     413&	18 35 00.4644&	0.0010&	-21 22 29.318&	0.013&	12.549&	0.020&	0.0046\\
29     417&	18 35 16.7906&	0.0010&	-21 19 04.158&	0.017&	13.003&	0.036&	0.0051\\
29     418&	18 35 25.4387&	0.0008&	-21 27 18.956&	0.014&	11.247&	0.020&	0.0043\\
29     420&	18 35 32.5528&	0.0006&	-21 27 11.170&	0.016&	12.163&	0.026&	0.0043\\
29     421&	18 35 37.8309&	0.0008&	-21 23 16.310&	0.015&	11.881&	0.019&	0.0043\\
29     422&	18 35 40.1481&	0.0010&	-21 17 27.260&	0.018&	12.525&	0.038&	0.0051\\
29     423&	18 35 41.8551&	0.0042&	-21 17 26.612&	0.018&	12.614&	0.037&	0.0050\\
\hline
\end{tabular}
\end{flushleft}
\end{table*}

\begin{table*}[tb]
\caption{Secondary catalog of references for the LU window (mean epoch 2450623, based on 24 nights
of observations)}
\label{tabrmm}
\begin{flushleft}
\begin{tabular}{@{}lcrcrccc@{}}
\hline
\rule{0pt}{1.2em}Name& $\alpha$ ($^{h}$ $^{m}$ $^{s}$)&$\sigma_{\alpha}$ ($^{s}$)&$\delta$ ($^{o}$ ' '')&$\sigma_{\delta} ('')$&$m_{Val}$&$\sigma_{m_{Val}}$&$\sigma_{w}$\\
\hline
30     238&	18 49 28.5675&	0.0082&	-13 16 22.926&	0.012&	12.606&	0.033&	0.0045\\
30     239&	18 49 31.9082&	0.0007&	-13 15 47.727&	0.012&	11.416&	0.021&	0.0060\\
30     241&	18 49 42.2207&	0.0021&	-13 14 05.520&	0.015&	12.955&	0.023&	0.0048\\
30     242 - T&	18 49 53.3125&	0.0015&	-13 14 33.781&	0.010&	10.547&	0.023&	0.0044\\
30     244&	18 49 54.6215&	0.0013&	-13 14 09.751&	0.012&	12.724&	0.026&	0.0042\\
30     249&	18 50 04.8668&	0.0009&	-13 22 17.704&	0.013&	12.297&	0.022&	0.0039\\
30     250&	18 50 06.5753&	0.0008&	-13 16 58.884&	0.018&	12.698&	0.038&	0.0054\\
30     251&	18 50 13.1341&	0.0008&	-13 14 23.728&	0.015&	12.754&	0.027&	0.0048\\
30     252 - T&	18 50 17.0313&	0.0012&	-13 15 27.649&	0.012&	10.105&	0.013&	0.0048\\
30     255 - T&	18 50 28.8516&	0.0008&	-13 18 24.981&	0.010&	11.503&	0.022&	0.0030\\
30     256&	18 50 29.8780&	0.0006&	-13 23 46.142&	0.012&	11.192&	0.017&	0.0061\\
30     258&	18 50 32.7738&	0.0012&	-13 15 02.493&	0.013&	12.867&	0.039&	0.0047\\
30     259&	18 50 33.6985&	0.0034&	-13 19 13.751&	0.012&	11.922&	0.053&	0.0040\\
30     260&	18 50 40.5369&	0.0011&	-13 19 22.359&	0.014&	12.836&	0.053&	0.0044\\
30     261&	18 50 46.9871&	0.0013&	-13 14 28.589&	0.010&	12.028&	0.024&	0.0044\\
30     262&	18 50 47.3931&	0.0008&	-13 22 22.143&	0.010&	12.014&	0.020&	0.0047\\
30     263&	18 50 49.8197&	0.0008&	-13 23 10.901&	0.010&	12.996&	0.024&	0.0047\\
30     264&	18 50 54.8421&	0.0010&	-13 17 41.822&	0.010&	11.981&	0.022&	0.0032\\
30     265&	18 50 56.2299&	0.0026&	-13 21 11.193&	0.013&	12.857&	0.035&	0.0049\\
30     266&	18 50 56.9783&	0.0015&	-13 20 16.203&	0.010&	12.188&	0.049&	0.0034\\
30     267&	18 51 00.4849&	0.0023&	-13 15 49.383&	0.014&	12.933&	0.044&	0.0050\\
30     268&	18 51 03.8975&	0.0019&	-13 16 16.112&	0.011&	12.330&	0.027&	0.0038\\
30     269&	18 51 19.3999&	0.0009&	-13 14 04.023&	0.013&	12.124&	0.028&	0.0036\\
30     271&	18 51 25.2372&	0.0008&	-13 19 43.627&	0.015&	12.853&	0.037&	0.0044\\
30     276&	18 51 47.3278&	0.0006&	-13 22 25.038&	0.011&	11.553&	0.021&	0.0035\\
30     278&	18 52 02.6973&	0.0010&	-13 17 18.220&	0.014&	12.113&	0.020&	0.0067\\
30     281&	18 52 22.0571&	0.0006&	-13 16 52.794&	0.012&	12.213&	0.026&	0.0040\\
30     282&	18 52 22.7781&	0.0006&	-13 14 48.627&	0.014&	12.977&	0.029&	0.0045\\
30     283&	18 52 23.6953&	0.0090&	-13 21 49.349&	0.017&	12.663&	0.030&	0.0044\\
30     284&	18 52 36.1103&	0.0019&	-13 13 41.872&	0.015&	11.498&	0.018&	0.0043\\
30     285&	18 52 36.6693&	0.0006&	-13 17 57.687&	0.014&	11.319&	0.019&	0.0046\\
\hline
\end{tabular}
\end{flushleft}
\end{table*}

\begin{table*}[tb]
\caption{Secondary catalog of references for the LV window (mean epoch 2450613, based on 23 nights
of observations)}
\label{tabrmm}
\begin{flushleft}
\begin{tabular}{@{}lcrcrccc@{}}
\hline
\rule{0pt}{1.2em}Name& $\alpha$ ($^{h}$ $^{m}$ $^{s}$)&$\sigma_{\alpha}$ ($^{s}$)&$\delta$ ($^{o}$ ' '')&$\sigma_{\delta} ('')$&$m_{Val}$&$\sigma_{m_{Val}}$&$\sigma_{w}$\\
\hline
31     157&	18 55 21.7316&	0.0009&	-10 20 48.162&	0.014&	12.538&	0.026&	0.0049\\
31     159&	18 55 31.1227&	0.0028&	-10 23 27.971&	0.013&	11.507&	0.029&	0.0051\\
31     161&	18 55 33.4437&	0.0015&	-10 21 20.079&	0.019&	12.437&	0.038&	0.0050\\
31     164&	18 55 38.8411&	0.0013&	-10 27 17.378&	0.015&	12.423&	0.026&	0.0044\\
31     166&	18 55 47.1937&	0.0008&	-10 22 30.979&	0.016&	12.820&	0.025&	0.0052\\
31     167&	18 55 50.1300&	0.0007&	-10 18 09.897&	0.015&	12.539&	0.043&	0.0050\\
31     169&	18 55 55.0718&	0.0006&	-10 19 36.954&	0.015&	11.732&	0.028&	0.0055\\
31     170&	18 56 00.5676&	0.0010&	-10 21 33.940&	0.016&	12.862&	0.030&	0.0052\\
31     176&	18 56 22.5299&	0.0009&	-10 19 08.881&	0.020&	12.559&	0.050&	0.0049\\
31     177&	18 56 29.4069&	0.0007&	-10 23 10.568&	0.015&	10.440&	0.040&	0.0068\\
31     178&	18 56 41.3249&	0.0008&	-10 21 10.318&	0.014&	12.414&	0.025&	0.0041\\
31     182&	18 56 55.0805&	0.0008&	-10 22 43.116&	0.020&	12.299&	0.024&	0.0054\\
31     184&	18 57 05.2617&	0.0012&	-10 26 40.166&	0.017&	12.555&	0.032&	0.0046\\
31     186&	18 57 07.2879&	0.0007&	-10 25 43.947&	0.019&	12.716&	0.060&	0.0049\\
31     187&	18 57 07.7382&	0.0014&	-10 21 57.860&	0.021&	12.496&	0.024&	0.0041\\
31     189&	18 57 10.3619&	0.0009&	-10 26 45.125&	0.020&	12.284&	0.035&	0.0044\\
31     190&	18 57 14.6889&	0.0007&	-10 23 20.256&	0.019&	10.962&	0.027&	0.0043\\
31     192&	18 57 17.8181&	0.0009&	-10 20 32.767&	0.022&	12.162&	0.025&	0.0048\\
31     193&	18 57 17.9920&	0.0018&	-10 18 36.855&	0.017&	12.993&	0.032&	0.0052\\
31     194&	18 57 22.3887&	0.0024&	-10 19 28.747&	0.019&	12.677&	0.029&	0.0049\\
31     195&	18 57 22.7461&	0.0023&	-10 23 19.257&	0.017&	13.001&	0.043&	0.0057\\
31     196&	18 57 25.7099&	0.0008&	-10 26 07.399&	0.019&	12.691&	0.042&	0.0045\\
31     197&	18 57 33.0289&	0.0059&	-10 23 06.122&	0.019&	11.853&	0.022&	0.0042\\
31     199&	18 57 40.5789&	0.0018&	-10 19 29.498&	0.015&	11.493&	0.023&	0.0036\\
31     200&	18 57 40.8931&	0.0011&	-10 22 25.342&	0.020&	12.991&	0.037&	0.0054\\
31     201&	18 57 43.5652&	0.0007&	-10 19 58.582&	0.014&	11.954&	0.023&	0.0038\\
31     202&	18 57 43.6110&	0.0017&	-10 20 56.706&	0.019&	12.950&	0.036&	0.0054\\
31     203 - T&	18 57 51.7813&	0.0008&	-10 22 02.922&	0.015&	11.227&	0.015&	0.0044\\
31     204&	18 57 56.4217&	0.0013&	-10 20 55.580&	0.015&	12.651&	0.028&	0.0044\\
31     205&	18 57 56.5643&	0.0009&	-10 17 39.462&	0.015&	11.564&	0.043&	0.0043\\
31     208&	18 58 08.8060&	0.0008&	-10 24 00.071&	0.011&	12.523&	0.037&	0.0042\\
31     209&	18 58 13.3357&	0.0022&	-10 18 23.395&	0.013&	12.712&	0.036&	0.0050\\
31     210&	18 58 17.7491&	0.0012&	-10 20 49.183&	0.015&	12.806&	0.025&	0.0060\\
31     215&	18 58 33.9723&	0.0017&	-10 23 16.148&	0.015&	12.685&	0.033&	0.0045\\
31     217&	18 58 53.1225&	0.0007&	-10 21 15.493&	0.013&	12.523&	0.024&	0.0047\\
31     220&	18 59 01.9055&	0.0010&	-10 26 33.559&	0.014&	12.342&	0.043&	0.0050\\
31     222&	18 59 13.2045&	0.0022&	-10 22 10.099&	0.010&	12.091&	0.037&	0.0042\\
31     223&	18 59 14.8563&	0.0021&	-10 22 48.533&	0.011&	12.407&	0.036&	0.0042\\
31     224&	18 59 16.4631&	0.0022&	-10 25 07.693&	0.012&	12.784&	0.048&	0.0052\\
31     225&	18 59 23.9396&	0.0008&	-10 19 49.779&	0.012&	12.479&	0.018&	0.0044\\
31     228&	18 59 32.4092&	0.0021&	-10 24 01.734&	0.011&	12.625&	0.043&	0.0048\\
31     229&	18 59 35.9876&	0.0018&	-10 26 06.872&	0.016&	12.775&	0.056&	0.0048\\
31     231&	18 59 39.0721&	0.0042&	-10 23 03.321&	0.016&	12.584&	0.038&	0.0046\\
31     233&	18 59 42.9261&	0.0014&	-10 23 35.108&	0.022&	12.172&	0.050&	0.0048\\
\hline
\end{tabular}
\end{flushleft}
\end{table*}


\begin{thebibliography}{}


\bibitem[1951]{baade} Baade W., 1951,
      Publ. Obs Univ. Mich., 10, 7

 \bibitem[1989]{paulo} Benevides-Soares P., Teixeira R., 1992,
      A\&A, 253, 307


\bibitem[1988]{blanco} Blanco V.M., 1988,
     AJ, 95, 1400.

   \bibitem[1989]{terndr} Blanco V.M., Terndrup D.M., 1989,
      AJ, 98, 843

   \bibitem[1998]{tania} Dominici T.P., 1998,
      MSc. Thesis IAG/USP

   \bibitem[1989]{eich} Eichhorn H., 1960,
      Astron. Nachr., 285, 233

   \bibitem[1997]{esa} ESA SP-1200, 1997,
      in: The Hipparcos and Tycho catalogues

   \bibitem[1988]{gcvs} Khopolov P.N., et al., 1988,
      in: General Catalog of Variable Stars,
      Four edition, Moscow "Nauka" Publish.

   \bibitem[1982]{nsv} Kukarkin B.V., et al., 1982,
      in: New Suspected Variable Stars,
      Moscow "Nauka" Publish.

   \bibitem[1996]{pac} Paczy\'nski B., 1996,
      Annu. Rev. Astron. Astrophys. 34, 419

   \bibitem[1989]{rama} Teixeira R., R\'equi\` eme Y., Benevides-Soares P.,  
 Rapaport M., 1992,
      A\&A, 264, 307

  \bibitem[1998]{bruno} Viateau, B. et al, 1998,
     accepted for publication in A\&AS.



 


\end{thebibliography}
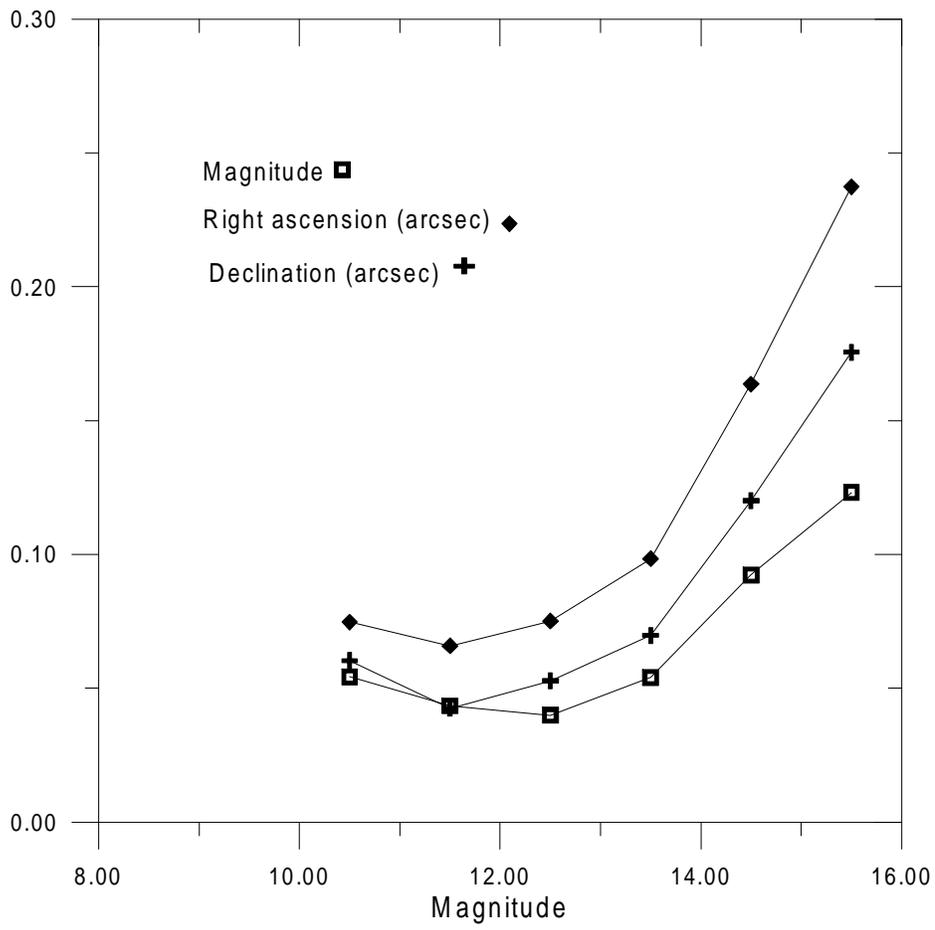
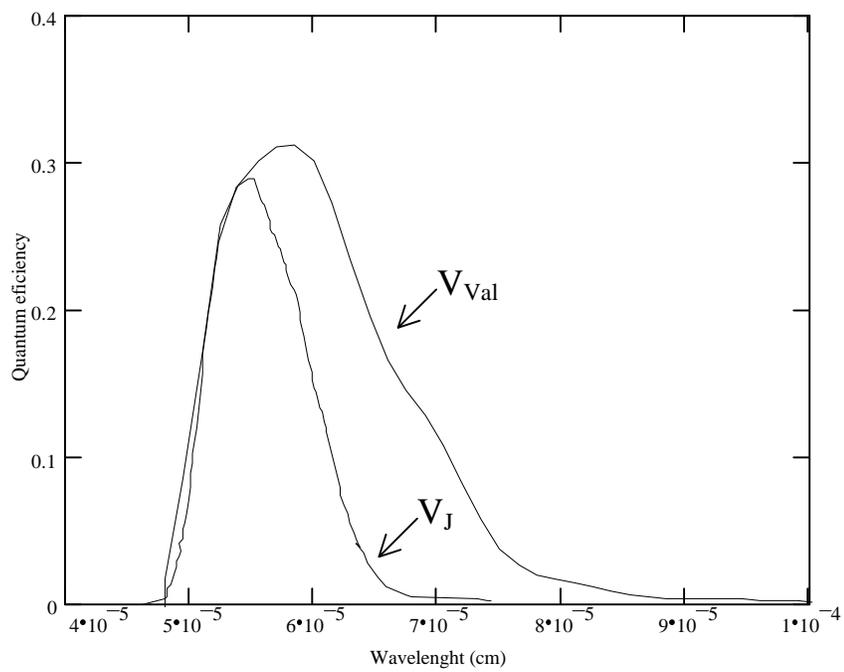
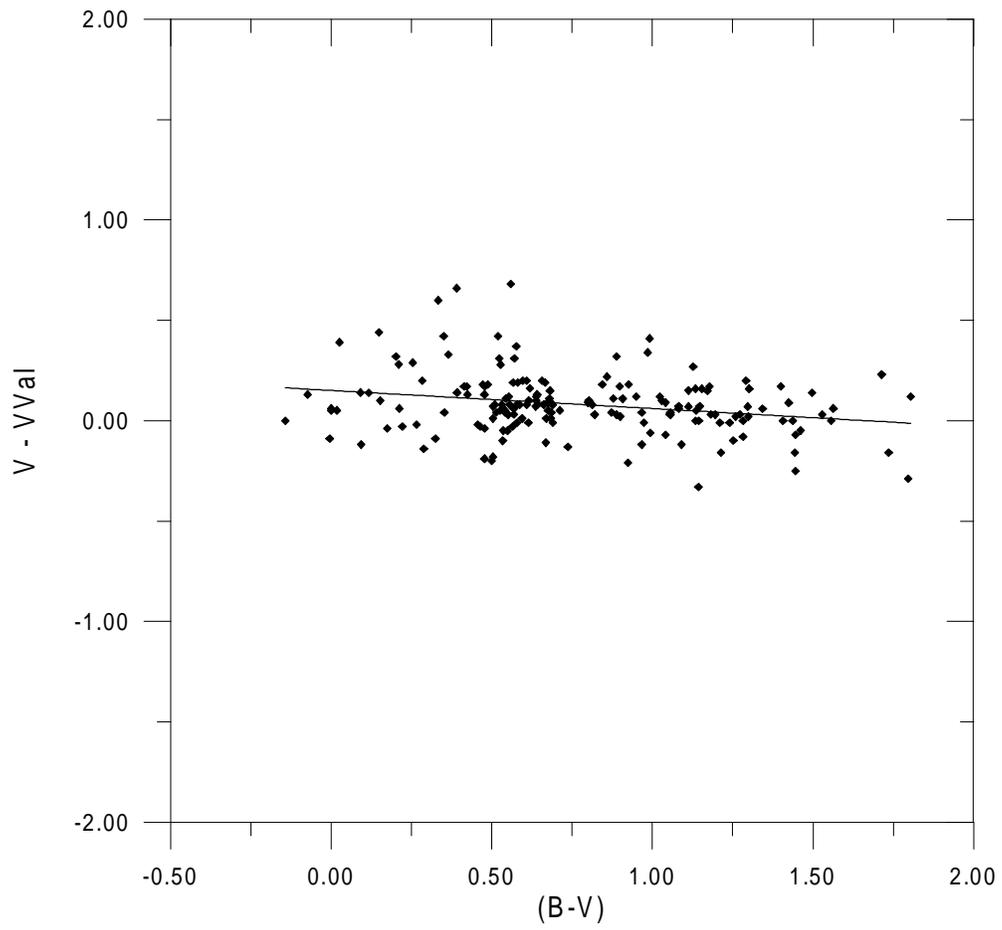
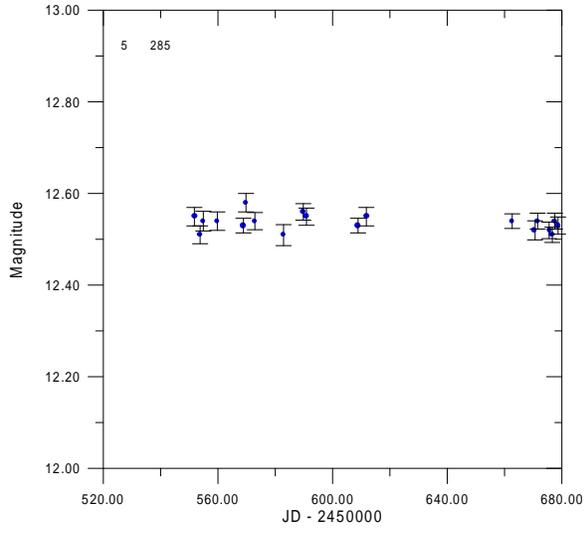
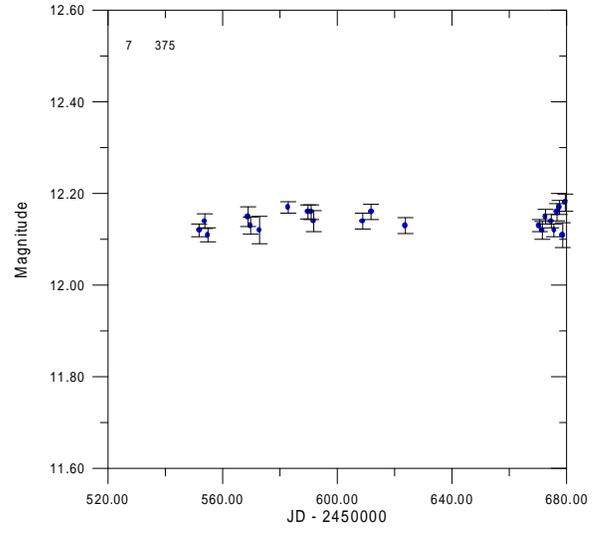
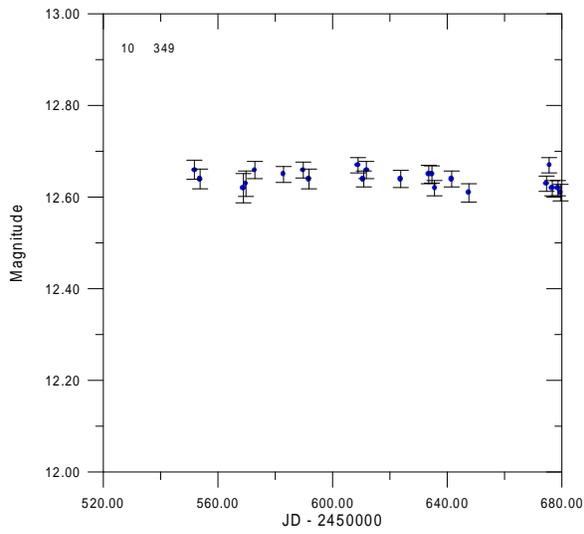
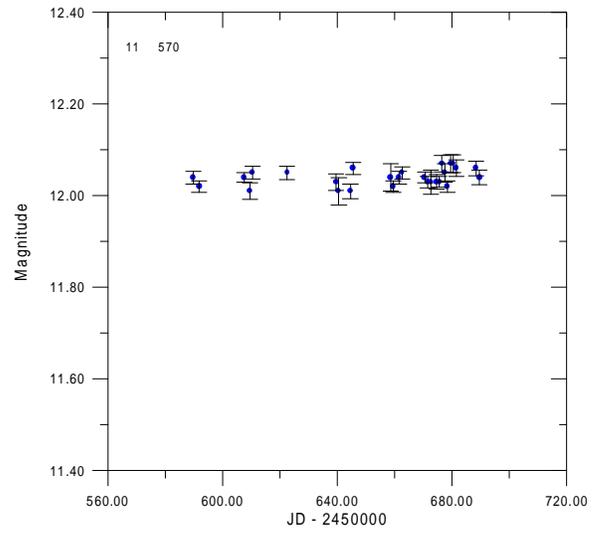
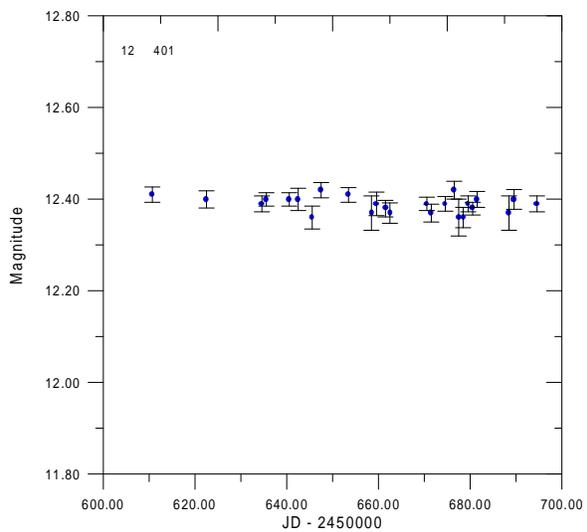
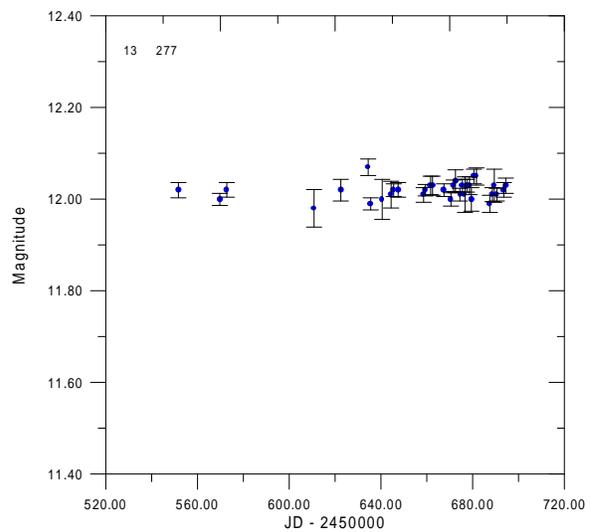
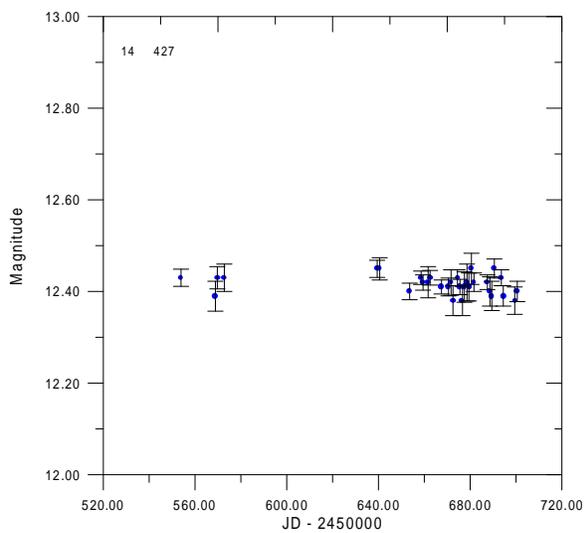
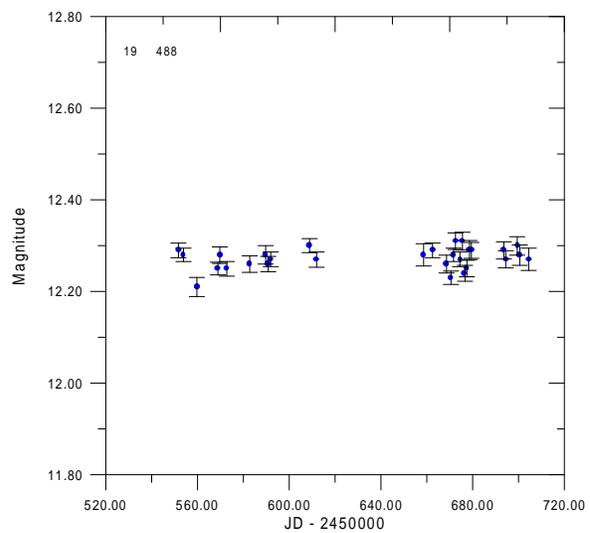
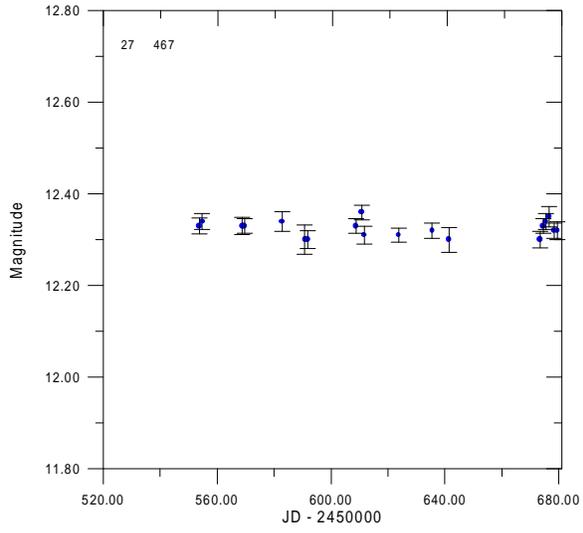
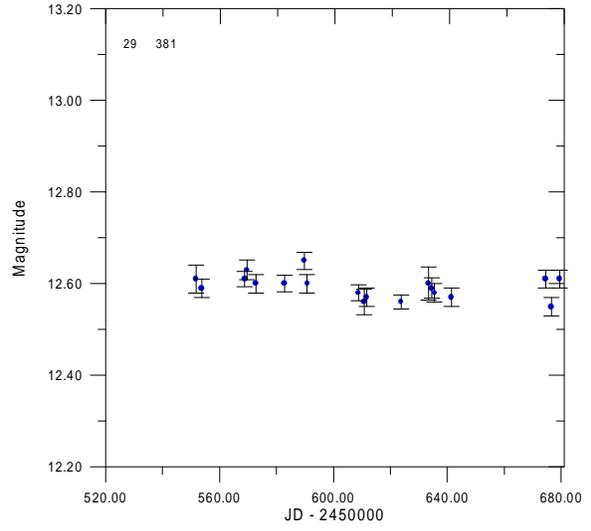
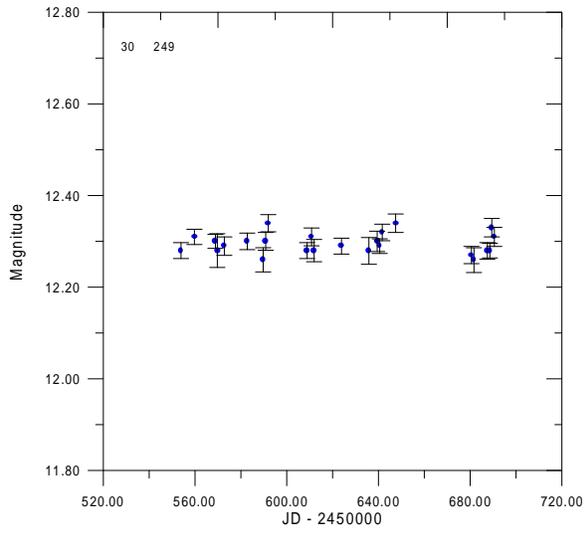
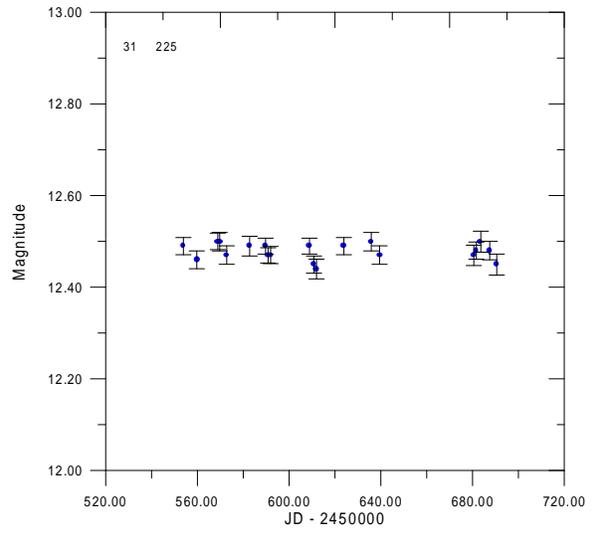
\end{document}